\documentclass[11pt, preprintnumbers,showkeys,showpacs,nofootinbib]{revtex4-2}

\usepackage{xcolor}

\usepackage[margin=1in]{geometry}
\usepackage{tabularx}
\usepackage{booktabs}
\usepackage{array}      

\usepackage{todonotes}
\usepackage{ragged2e}
\usepackage{graphicx,overpic}
\usepackage{latexsym}
\usepackage{amsmath}
\usepackage{amsfonts}
\usepackage{amssymb}

\usepackage{verbatim}
\usepackage{dcolumn}
\usepackage{bm}
\usepackage{float}
\usepackage{subfigure}
\usepackage[english]{babel}

\usepackage[colorlinks=true, pdfstartview=FitV, linkcolor=blue, citecolor=red, urlcolor=magenta]{hyperref}
\usepackage{tikz}
\usepackage[export]{adjustbox}

\usepackage{csquotes}
\usepackage{soul}

\begin{document}
\newcommand{\be}{\begin{equation}}
\newcommand{\ee}{\end{equation}}
\newcommand{\bea}{\begin{eqnarray}}
\newcommand{\eea}{\end{eqnarray}}
\newcommand{\hf}{\frac{1}{2}}
\newcommand{\pa}{\partial}

\newcommand{\NITK}{
\affiliation{Department of Physics, National Institute of Technology Karnataka, Surathkal  575 025, India}
}

\newcommand{\UW}{\affiliation {Department of Oral Health Sciences, School of Dentistry, University of Washington, Seattle, WA 98195, USA}}

\newcommand{\naveencroatia}{\affiliation {Theoretical Physics Division, Rudjer Bo\v skovi\'c Institute,\\  Bijeni\v cka c.54,  HR-10002 Zagreb, Croatia}}

\title{All-Order Linearized Hydrodynamics of Charged Holographic Plasmas }
\author{Karthik R. }
\email{raokarthik1198@gmail.com, karthikr.227ph003@nitk.edu.in}
\NITK

\author{Shamant Shreedhar Dixit}
\email{dixit.shamant@gmail.com}
\NITK

\author{K. M. Ajith}
\email{ajith@nitk.edu.in}
\NITK

\author{Kartheek Hegde}
\email{hegde.kartheek@gmail.com}
\NITK

\author{Shreyas Punacha}
\email{shreyas4@uw.edu}
\UW

\author{A. Naveena Kumara}
\email{naviphysics@gmail.com, nathith@irb.hr}
\naveencroatia

\begin{abstract}
We investigate linearized all-order transport in a finite-density holographic plasma described by five-dimensional Einstein-Maxwell-Chern-Simons theory. By retaining the full frequency- and momentum-dependence of the fluid/gravity equations, we determine generalized shear, diffusion, higher-gradient, and chiral-vortical transport functions together with their real-time memory kernels. The different channels exhibit a clear hierarchy of temporal response: the leading shear and diffusion sectors respond promptly, whereas the higher-gradient sectors develop only after a finite buildup time. Remarkably, the chiral-vortical response also shows a finite buildup time despite entering at first order in the hydrodynamic derivative expansion, demonstrating that derivative order does not by itself determine the temporal organization of transport. Toward the near-extremal regime, long-time memory becomes increasingly persistent across several channels. These results reveal how finite density reshapes nonlocal transport and relaxation in strongly coupled holographic plasmas.
\end{abstract}



\maketitle

 \section{Introduction}

Hydrodynamics provides an effective description of interacting many-body systems at sufficiently long wavelengths and low frequencies, where the dynamics is controlled by conserved quantities and their associated collective modes~\cite{landau1987fluid,forster1975hydrodynamic,Rezzolla:2013dea}. The macroscopic evolution is encoded in conservation laws for the energy-momentum tensor and conserved currents, supplemented by constitutive relations expressed in terms of local hydrodynamic variables such as the temperature, fluid velocity, and charge density. Near local equilibrium, these constitutive relations are organized systematically in a derivative expansion, with transport coefficients characterizing the response of the underlying microscopic theory~\cite{Romatschke:2017vte,Kovtun:2012rj}. Determining these coefficients is particularly challenging in strongly coupled systems, where conventional perturbative methods are generally unavailable.

The derivative expansion, however, has an intrinsically limited regime of applicability. At first order, relativistic Navier-Stokes theory leads to acausal and unstable evolution equations~\cite{Hiscock:1983zz,Hiscock:1985zz,Hiscock:1987zz}, motivating causal extensions such as Israel-Stewart theory~\cite{Israel:1976tn,Israel:1979wp} and, more recently, alternative reorganizations of relativistic hydrodynamics~\cite{Bemfica:2019knx,Kovtun:2019hdm}. More generally, finite-order truncations become inadequate once frequencies or momenta approach microscopic relaxation scales. In many interacting systems the hydrodynamic gradient expansion is asymptotic, with its large-order behavior tied to non-hydrodynamic excitations~\cite{Heller:2013fn,Heller:2015dha,Aniceto:2018bis,Grozdanov:2019kge}. These observations motivate formulations that retain the hydrodynamic variables while incorporating their response beyond any fixed order in gradients.

Gauge/gravity duality provides a controlled setting in which such questions can be studied at strong coupling~\cite{Maldacena:1997re,Gubser:1998bc,Witten:1998qj}. Thermal states of the boundary theory are described by asymptotically anti-de Sitter black branes, while long-wavelength fluctuations of these geometries reproduce the hydrodynamic dynamics of the dual plasma. This relation is made systematic by the fluid/gravity correspondence~\cite{Bhattacharyya:2007vjd,Rangamani:2009xk}, in which slowly varying black-brane parameters are promoted to boundary fields and the bulk equations are solved order by order in derivatives. The resulting gravitational solution determines the boundary constitutive relations and transport coefficients. The construction was originally developed for neutral conformal fluids and was subsequently extended to charged black branes, where both the stress tensor and conserved current participate in the hydrodynamic dynamics~\cite{Erdmenger:2008rm,Banerjee:2008th}.

An important extension of this framework is obtained by linearizing in the amplitude of the hydrodynamic perturbations while retaining their dependence on frequency and spatial momentum to all orders. In this formulation, the infinite derivative expansion is resummed into generalized transport functions, such as a frequency- and momentum-dependent shear viscosity, rather than represented by an ever-growing set of individual transport coefficients~\cite{Lublinsky:2009kv,Bu:2014ena}. The corresponding constitutive relations are therefore nonlocal in spacetime while remaining linear in the perturbations. In the time domain, this nonlocality is naturally expressed through memory kernels, which characterize how the present response of the plasma depends on its previous evolution and provide a direct way of examining causality and relaxation. All-order constructions of this type have been developed for neutral holographic fluids and subsequently generalized to conserved probe currents and higher-derivative gravitational theories~\cite{Bu:2014ena,Bu:2015ika,Bu:2015ame,Bu:2015bwa,Bu:2020jfo}.

Finite density introduces an additional layer of dynamics because fluctuations of the metric and gauge field are coupled through the bulk equations. This coupling allows momentum transport, charge diffusion, and, in the presence of a Chern-Simons interaction, anomaly-induced transport to be treated within a common framework~\cite{Erdmenger:2008rm,Banerjee:2008th,Son:2009tf}. It is also of broader interest because transport at finite chemical potential is relevant to strongly interacting matter in regimes where first-principles real-time calculations are particularly difficult~\cite{Achenbach:2023pba,Bzdak:2019pkr,Meyer:2011gj,Aarts:2015tyj}. Despite considerable progress in charged fluid/gravity and anomalous hydrodynamics~\cite{Erdmenger:2008rm,Banerjee:2008th,Son:2009tf,Bu:2016oba,Bu:2016vum,Landsteiner:2016led}, the all-order linear response of a fully backreacted Einstein-Maxwell-Chern-Simons plasma has received much less attention than its neutral or probe-field counterparts. In particular, the simultaneous frequency- and momentum-dependent behavior of viscous, diffusive, and anomaly-induced transport, together with their real-time memory kernels, remains to be systematically characterized.

In this work, we develop the all-order linearized fluid/gravity construction for a charged holographic plasma described by five-dimensional Einstein-Maxwell-Chern-Simons theory. We first solve the bulk equations perturbatively in the hydrodynamic expansion, recovering the known leading-order shear, diffusion, and chiral-vortical transport coefficients and obtaining higher-order derivative corrections. This provides analytical control of the hydrodynamic limit and a benchmark for the all-order calculation. We then solve the coupled radial equations numerically using a Chebyshev pseudospectral method without truncating the frequency and momentum dependence, thereby determining the generalized transport functions across the hydrodynamic and non-hydrodynamic regimes. Their inverse Fourier transforms provide the corresponding memory kernels and allow us to study directly how finite density modifies the temporal organization of transport.

A central result of our analysis is that the various transport channels display qualitatively different real-time responses. The generalized shear and diffusion kernels exhibit a prompt response, whereas the higher-gradient shear and charge-gradient tensor channels develop only after a finite buildup time. Notably, the generalized chiral-vortical response also exhibits a finite buildup time despite contributing already at first order in the hydrodynamic expansion, showing that the derivative order of a transport term does not by itself determine its temporal response. Increasing the charge parameter suppresses the ordinary charge-diffusion coefficient while enhancing the persistence of several memory kernels, with relaxation becoming increasingly long-lived toward the near-extremal regime. The latter behavior is suggestive of the increasingly important infrared dynamics associated with the near-horizon geometry of nearly extremal charged black branes~\cite{Berti:2003ud,Faulkner:2009wj,Edalati:2009bi,Edalati:2010hk,Davison:2013bxa,He:2021jna}.

The remainder of the paper is organized as follows. In Sec.~\ref{section-2}, we introduce the charged black-brane background and formulate the linearized fluid/gravity problem. In Sec.~\ref{section-3}, we develop the perturbative hydrodynamic expansion and extract the corresponding transport coefficients. Section~\ref{section-4} presents the all-order numerical construction. In Sec.~\ref{section-5}, we analyze the resulting transport functions and their memory kernels, including the shear, diffusion, higher-gradient, and anomalous sectors. We summarize the physical picture and discuss open questions in Sec.~\ref{section-6}.

\section{Holographic setup and linearized fluid/gravity}
\label{section-2}

In this section, we formulate the holographic problem underlying the all-order transport analysis. We first introduce the charged AdS$_5$ black-brane background and its thermodynamic parametrization, and then construct linearized perturbations around this equilibrium state. After specifying the gauge, boundary conditions, and hydrodynamic frame, we decompose the bulk equations into independent scalar, vector, and tensor structures. The resulting radial equations determine the frequency- and momentum-dependent transport functions studied in the subsequent sections.

\subsection{Einstein-Maxwell-Chern-Simons black brane}

We consider five-dimensional Einstein gravity coupled to a $U(1)$ gauge field with a Chern-Simons interaction, defined on an asymptotically $\mathrm{AdS}_5$ spacetime~\cite{Banerjee:2008th}. Our index conventions are as follows: uppercase Latin indices $A,B\in\{v,r,x,y,z\}$ denote bulk coordinates, Greek indices $\mu,\nu\in\{v,x,y,z\}$ denote boundary coordinates, and lowercase Latin indices $i,j\in\{x,y,z\}$ denote spatial boundary directions. The action is
\begin{equation}
S=
\frac{1}{16\pi G_5}
\int d^5x\,\sqrt{-g_5}
\left[
R+12-F_{AB}F^{AB}
-\frac{4\kappa}{3}
\epsilon^{LABCD}A_LF_{AB}F_{CD}
\right].
\end{equation}
For the supersymmetric value $\kappa=1/(2\sqrt{3})$, this theory arises as a consistent truncation of type IIB supergravity on $\mathrm{AdS}_5\times S^5$. We nevertheless keep $\kappa$ arbitrary in the derivation and specialize to the supersymmetric value only when presenting the numerical results.

The equations of motion are
\begin{equation}
\label{eq:2}
\begin{aligned}
&G_{AB}-6g_{AB}
+2\left(
F_{AC}F^{C}{}_{B}
+\frac{1}{4}g_{AB}F_{CD}F^{CD}
\right)=0,
\\
&\nabla_BF^{AB}
+\kappa\epsilon^{ABCDE}F_{BC}F_{DE}=0.
\end{aligned}
\end{equation}
For later convenience, we denote the left-hand sides of the Einstein and Maxwell equations by $E_{AB}$ and $\mathcal{M}^{A}$, respectively.

Equations~\eqref{eq:2} admit the charged AdS-Reissner-Nordstr\"om black-brane solution~\cite{Chamblin:1999tk,Erdmenger:2008rm,Banerjee:2008th}
\begin{equation}
\label{eq:3}
\begin{aligned}
ds^2 &=
-2u_\mu dx^\mu dr
-r^2f(r,m,q)\,u_\mu u_\nu dx^\mu dx^\nu
+r^2\mathcal{P}_{\mu\nu}dx^\mu dx^\nu,
\\
A &=
\frac{\sqrt{3}q}{2r^2}\,u_\mu dx^\mu ,
\end{aligned}
\end{equation}
where
\begin{equation}
\begin{aligned}
f(r,m,q)
&=
1-\frac{m}{r^4}+\frac{q^2}{r^6},
\\
\mathcal{P}_{\mu\nu}
&\equiv
\eta_{\mu\nu}+u_\mu u_\nu,
\qquad
\eta_{\mu\nu}
=
\mathrm{diag}(-1,1,1,1).
\end{aligned}
\end{equation}
For a uniformly boosted brane,
\begin{equation}
u_v=-\frac{1}{\sqrt{1-\beta^2}},
\qquad
u_i=\frac{\beta_i}{\sqrt{1-\beta^2}},
\qquad
\beta^2=\sum_{i=1}^{3}\beta_i\beta^i .
\end{equation}
The constant parameters $m$, $q$, and $u_\mu$ specify a family of equilibrium black-brane solutions and, holographically, equilibrium states of the dual charged plasma.

Following~\cite{Banerjee:2008th}, we introduce the dimensionless variables~\footnote{Unless explicitly rescaled, $r,m,q,\omega$, and spatial momentum are dimensionful. We introduce $\rho=r/R$, $M=m/R^4$, and $Q=q/R^3$ when writing dimensionless equations.}
\begin{equation}
\label{eq:6}
\rho\equiv\frac{r}{R},
\qquad
M\equiv\frac{m}{R^4},
\qquad
Q\equiv\frac{q}{R^3},
\qquad
Q^2=M-1,
\end{equation}
where $R$ denotes the outer-horizon radius. The Hawking temperature, chemical potential, and charge density are
\begin{equation}
\label{eq:7}
T=
\frac{R}{2\pi}(2-Q^2),
\qquad
\mu=
\frac{2\sqrt{3}q}{R^2}=2\sqrt{3}QR,
\qquad
n=
\frac{\sqrt{3}q}{16\pi G_5}.
\end{equation}
The outer and inner horizons are located at
\begin{equation}
\rho_+=1,
\qquad
\rho_-=
\left[
\left(Q^2+\frac14\right)^{1/2}
-\frac12
\right]^{1/2}.
\end{equation}
Extremality corresponds to $\rho_+=\rho_-$, or equivalently $Q^2=2$ and $M=3$. At this point the horizon becomes degenerate and the radial problem requires a separate treatment. We therefore restrict the analysis to
\begin{equation}
0\leq Q^2<2,
\qquad
1\leq M<3,
\end{equation}
We include near-extremal configurations, including $M=2.99$, and check numerical convergence as the horizon approaches degeneracy.

\subsection{Linearized fluid/gravity ansatz}

The uniform solution~\eqref{eq:3} describes an equilibrium state of the boundary theory. Hydrodynamic configurations are generated by allowing the local velocity, mass parameter, and charge parameter to vary over the boundary coordinates. The resulting locally boosted black-brane geometry may be regarded as a local-equilibrium seed solution, but it no longer satisfies the bulk equations once derivatives of the boundary fields are included. The required bulk corrections encode the dissipative and higher-gradient response of the dual fluid~\cite{Bhattacharyya:2007vjd,
Banerjee:2008th}.

The corresponding seed fields are
\begin{equation}
\label{eq:10}
\begin{aligned}
ds^2={}&
-2u_\mu(x)\,dx^\mu dr
-r^2
f\bigl(r,m(x),q(x)\bigr)
u_\mu(x)u_\nu(x)\,dx^\mu dx^\nu
+r^2\mathcal{P}_{\mu\nu}(x)\,dx^\mu dx^\nu,
\\
A_M^{(0)}dx^M={}&
\frac{\sqrt{3}\,q(x)}{2r^2}
u_\mu(x)\,dx^\mu .
\end{aligned}
\end{equation}
In conventional fluid/gravity, the corrections to~\eqref{eq:10} are constructed order by order in boundary derivatives. Here our objective is instead to retain the derivative dependence to all orders while working only to linear order in the amplitude of the hydrodynamic perturbations, following the linearized fluid/gravity strategy of Ref.~\cite{Bu:2014ena}.

We therefore expand around a homogeneous equilibrium state,
\begin{equation}
\label{eq:11}
\begin{aligned}
m(x)
&=
m_0+\epsilon\,m^{(0)}(x)
+\mathcal{O}(\epsilon^2),
\\
q(x)
&=
q_0+\epsilon\,q^{(0)}(x)
+\mathcal{O}(\epsilon^2),
\\
u_\mu(x)
&=
\left(-1,\epsilon\,\beta_i(x)\right)
+\mathcal{O}(\epsilon^2),
\end{aligned}
\end{equation}
where $\epsilon$ is an amplitude-counting parameter. We retain terms through $\mathcal{O}(\epsilon)$ and set $\epsilon=1$ after linearization.

Substituting~\eqref{eq:11} into the seed solution gives
\begin{equation}
\label{eq:12}
\begin{aligned}
ds^2={}&
2\,dv\,dr
-r^2f^{(0)}(r)\,dv^2
+r^2dx_i dx^i
\\
&-\epsilon
\Bigg[
2\beta_i(x)\,dx^i dr
+2\beta_i(x)r^2
\left(1-f^{(0)}(r)\right)dx^i dv
+\left(
-\frac{m^{(0)}(x)}{r^2}
+\frac{2q_0q^{(0)}(x)}{r^4}
\right)dv^2
\Bigg]
+\mathcal{O}(\epsilon^2),
\end{aligned}
\end{equation}
with
\begin{equation}
f^{(0)}(r)
=
1-\frac{m_0}{r^4}
+\frac{q_0^2}{r^6},
\end{equation}
while the gauge field becomes
\begin{equation}
\label{eq:14}
A=
-\frac{\sqrt{3}}{2}
\left[
\frac{q_0+\epsilon q^{(0)}(x)}{r^2}\,dv
-\epsilon\frac{q_0}{r^2}\beta_i(x)\,dx^i
\right]
+\mathcal{O}(\epsilon^2).
\end{equation}

Because the boundary fields now depend on $x^\mu$, the fields \eqref{eq:12} and~\eqref{eq:14} do not by themselves satisfy the bulk equations at $\mathcal{O}(\epsilon)$. We therefore write
\begin{equation}
g
=
g^{(0)}
+\epsilon
\left(
g_{\mathrm{in}}^{(1)}
+
g_{\mathrm{corr}}^{(1)}
\right)
+\mathcal{O}(\epsilon^2),
\end{equation}
and
\begin{equation}
A
=
A^{(0)}
+\epsilon
\left(
A_{\mathrm{in}}^{(1)}
+
A_{\mathrm{corr}}^{(1)}
\right)
+\mathcal{O}(\epsilon^2),
\end{equation}
where the ``in'' terms are generated directly by~\eqref{eq:11}, whereas the correction fields are determined by the bulk equations.

We adopt the standard fluid/gravity gauge ~\cite{Bhattacharyya:2007vjd,Banerjee:2008th},
\begin{equation}
\label{eq:17}
g_{rr}=0,
\qquad
g_{r\mu}\propto u_\mu,
\qquad
A_r=0,
\qquad
\mathrm{Tr}
\left[
(g^{(0)})^{-1}g_{\mathrm{corr}}^{(1)}
\right]
=0 .
\end{equation}
At linear order the correction fields can then be parameterized as
\begin{equation}
\begin{aligned}
ds^2_{\mathrm{corr}}
=
\epsilon
\left[
-3h\,dv\,dr
+r^2h\,d\vec{x}^{\,2}
+\frac{k}{r^2}dv^2
+12r^2j_i\,dv\,dx^i
+r^2\alpha_{ij}dx^i dx^j
\right],
\end{aligned}
\end{equation}
and
\begin{equation}
A_{\mathrm{corr}}^{(1)}
=
\frac{\sqrt{3}}{2}\epsilon
\left(
-\frac{w}{r^2}\,dv
+r^5g_i\,dx^i
\right).
\end{equation}
The scalar sector is described by $h$, $k$, and $w$, the vector sector by $j_i$ and $g_i$, and the symmetric traceless tensor sector by $\alpha_{ij}$.

\subsection{Boundary data, boundary conditions, and hydrodynamic frame}

The asymptotic behavior of the bulk fields determines the stress tensor and conserved current of the dual theory. Using holographic renormalization ~\cite{Gubser:1998bc,Witten:1998qj}, the renormalized stress tensor may be written in the Brown-York form
\begin{equation}
\label{eq:23}
T^{\mu}{}_{\nu}
=
-\frac{1}{8\pi G_5}\lim_{r\rightarrow\infty}
r^4
\left(
\mathcal{K}^{\mu}{}_{\nu}
-\mathcal{K}\gamma^{\mu}{}_{\nu}
+3\gamma^{\mu}{}_{\nu}
-\frac12 G^{\mu}{}_{\nu}[\gamma]
\right),
\end{equation}
where $\gamma_{\mu\nu}$ is the induced metric on a constant-$r$ hypersurface and $\mathcal{K}_{\mu\nu}$ its extrinsic curvature. In the conventions used here, the boundary current is extracted from the asymptotic gauge field as
\begin{equation}
\label{eq:24}
J_\mu
=
\lim_{r\rightarrow\infty}
\frac{r^2A_\mu}{8\pi G_5}.
\end{equation}
The equilibrium black-brane background generates the ideal-fluid contributions, whereas the normalizable parts of the correction fields encode dissipative and higher-gradient transport.

At leading hydrodynamic order the current has the schematic form
\begin{equation}
J_\mu
=
nu_\mu
-D\,
\left(\mathcal{P}_{\mu}{}^{\nu}\nabla_\nu n+3na_\mu\right)
+\xi\,\ell_\mu
+\mathcal{O}(\pa^2),\qquad a_\mu=u^{\nu}\nabla_\nu u_\mu.
\end{equation}
where
\begin{equation}
\ell_\mu
=
\epsilon_{\mu\nu\rho\sigma}
u^\nu\nabla^\rho u^\sigma
\end{equation}
is the vorticity pseudovector. The coefficient $D$ controls charge diffusion, while $\xi$ is generated by the Chern-Simons coupling and describes the chiral-vortical response~\cite{Son:2009tf, Erdmenger:2008rm,Banerjee:2008th}.

The radial equations are solved subject to three physical requirements. First, all bulk perturbations must be regular at the future horizon. In ingoing Eddington-Finkelstein coordinates, the physically relevant ingoing solution is the solution regular at the horizon. Second, the boundary geometry is held fixed, so non-normalizable deformations of the boundary metric and gauge source are excluded. The corresponding large-$r$ behavior is
\begin{equation}
\label{eq:27}
h<\mathcal{O}(r^0),
\qquad
k<\mathcal{O}(r^4),
\qquad
w<\mathcal{O}(r^0),
\qquad
j_i<\mathcal{O}(r^0),
\qquad
g_i<\mathcal{O}(r^{-5}),
\qquad
\alpha_{ij}<\mathcal{O}(r^0).
\end{equation}
Finally, we work in the Landau frame,
\begin{equation}
\label{eq:28}
u_\mu T_{\mathrm{Diss}}^{\mu\nu}=0,\qquad u_\mu J^\mu_\mathrm{Diss}=0.
\end{equation}
These conditions define the local fluid velocity and thermodynamic variables so that the non-ideal corrections carry neither a local-rest-frame energy-density shift nor a charge-density shift.

\subsection{All-order radial equations}

The linearized Einstein-Maxwell-Chern-Simons equations naturally divide into dynamical and constraint equations. The dynamical equations determine the radial profiles of the correction fields, while the constraint equations reproduce the conservation laws of the dual boundary theory. This separation allows the transport functions to be determined without first imposing the hydrodynamic equations of motion on the boundary fields, as in the standard fluid/gravity construction ~\cite{Bhattacharyya:2007vjd,Banerjee:2008th,Bu:2014ena}.

The simplest scalar equation is $E_{rr}=0$, which gives
\begin{equation}
5\partial_r h+r\partial_r^2h=0.
\end{equation}
Its general solution is
\begin{equation}
h(r,x)
=
\frac{C_h^{(1)}(x)}{r^4}
+
C_h^{(2)}(x).
\end{equation}
The non-normalizable mode $C_h^{(2)}$ is eliminated by the boundary condition, while the Landau-frame condition removes $C_h^{(1)}$. Consequently,
\begin{equation}
h=0.
\end{equation}

The remaining scalar, vector, and tensor equations form a coupled linear system for $w$, $k$, $j_i$, $g_i$, and $\alpha_{ij}$. Rather than solving the component equations directly, we exploit spatial $SO(3)$ symmetry and linearity to decompose these fields into independent structures built from $\beta_i$ and $q^{(0)}$. For completeness, the explicit component form of the linearized Einstein-Maxwell-Chern-Simons equations prior to this decomposition is collected in Appendix.~\ref{app:linearized-bulk-equations}. We simultaneously Fourier transform the boundary coordinates according to
\begin{equation}
\partial_\mu\longrightarrow(-i\omega,i\vec{q}),
\end{equation}
so that the original partial differential equations reduce to ordinary differential equations in the radial coordinate.

Using the dimensionless variables introduced in~\eqref{eq:6}, the most general decomposition required for the linearized problem is
\begin{equation}
\label{eq:37}
\begin{aligned}
w
&=
a_1(\omega,q,\rho)\,q^{(0)}
+
a_2(\omega,q,\rho)\,
\partial_m\beta_m,
\\
j_i
&=
b_1(\omega,q,\rho)\,\beta_i
+
b_2(\omega,q,\rho)\,
\partial_i\partial_m\beta_m
+
b_3(\omega,q,\rho)\,
\partial_iq^{(0)}
+
b_4(\omega,q,\rho)\,
\epsilon_{ijk}\partial_j\beta_k,
\\
g_i
&=
V_1(\omega,q,\rho)\,\beta_i
+
V_2(\omega,q,\rho)\,
\partial_i\partial_m\beta_m
+
V_3(\omega,q,\rho)\,
\partial_iq^{(0)}
+
V_4(\omega,q,\rho)\,
\epsilon_{ijk}\partial_j\beta_k,
\\
\alpha_{ij}
&=
2s_1(\omega,q,\rho)\sigma_{ij}
+
s_2(\omega,q,\rho)\pi_{ij}
+
s_3(\omega,q,\rho)\Xi_{ij}+2s_4(\omega,q,\rho)\Omega_{ij},
\end{aligned}
\end{equation}
where
\begin{equation}
\label{eq:38}
\begin{aligned}
\sigma_{ij}
&=
\partial_{\langle i}\beta_{j\rangle}
=
\frac12
\left(
\partial_i\beta_j+\partial_j\beta_i
\right)
-\frac13\delta_{ij}\partial_m\beta_m,
\\
\Xi_{ij}
&=
\partial_{\langle i}\partial_{j\rangle}q^{(0)}
=
\left(
\partial_i\partial_j
-\frac13\delta_{ij}\partial^2
\right)q^{(0)},
\\
\pi_{ij}
&=
\partial_{\langle i}\partial_{j\rangle}
(\partial_m\beta_m)
=
\left(
\partial_i\partial_j
-\frac13\delta_{ij}\partial^2
\right)
(\partial_m\beta_m)\\
    \Omega_{ij}& = \frac{1}{2}\left(\partial_i\ell_j+\partial_j\ell_i\right).
\end{aligned}
\end{equation}

Homogeneous contributions associated with residual redefinitions of the hydrodynamic variables are fixed by horizon regularity, boundary normalizability, and the Landau-frame condition. Additional structures involving derivatives of $q^{(0)}$ are related to the basis above by the constraint equations and therefore do not introduce independent transport functions.

Substitution of~\eqref{eq:37} into the bulk equations yields five coupled radial subsystems. 

\par\medskip
\paragraph{\bf Sector I: $\{b_1,V_1,s_1\}$.}
\begin{equation}
\label{eq:39}
\begin{aligned}
0 {}= & \ 6r^4 \partial_r^2 b_1 + 30r^3 \partial_r b_1 + 6r^4 q_0 \partial_r V_1 + 30r^3 q_0 V_1 - q^2 r^2 \partial_r s_1 - q^2 - 3i\omega r, \\[1mm] 0= & \
        \left(q_0^2r^2-m_0 r^4+r^8\right)\partial_r^2 V_1 + \left(7q_0^2r-9m_0r^3+13r^7\right)\partial_r V_1 + 5\left(q_0^2-3m_0r^2+7r^6\right)V_1\\ & + 12q_0\partial_r b_1 - 11i\omega r^5 V_1 - 2i\omega r^6\partial_r V_1 - q^2r^4 V_1 - q^2 \frac{q_0}{r^3}+8\sqrt{3}\kappa q_0 r^2 q^2 V_4 + \frac{i\omega q_0}{r^2},  \\[1mm] 0= & \
        r\left(r^6-m_0r^2+q_0^2\right)\partial_r^2 s_1 - \left(-5r^6+m_0r^2+q_0^2\right)\partial_r s_1 - 2i\omega r^5\partial_r s_1 - 3i\omega r^4 s_1 - 18r^4 b_1 \\ & - 6r^5\partial_r b_1 + 3r^4 - i\omega r^3.
\end{aligned}
\end{equation}

\par\medskip
\paragraph{\bf Sector II: $\{b_2,V_2,s_2\}$.}
\begin{equation}
\label{eq:40}
\begin{aligned}
0{}= & \ 6r^4 \partial_r^2b_2 + 30r^3\partial_r b_2 + 6r^4 q_0\partial_r V_2 + 30r^3 q_0 V_2 + \frac{1}{3}r^2 \partial_r s_1 - \frac{2}{3}r^2q^2 \partial_r s_2 - 1,  \\[1mm] 0 = & \
        \left(q_0^2r^2-m_0r^4+r^8\right)\partial_r^2 V_2 + \left(7q_0^2 r-9m_0r^3+13r^7\right)\partial_r V_2 + 5\left(q_0^2-3m_0r^2+7r^6\right)V_2 \\ & + 12q_0\partial_r b_2 - 11i\omega r^5 V_2 - 2i\omega r^6\partial_r V_2 - r^4V_1+8\sqrt{3}\kappa q_0 r^2 V_4  + \frac{1}{r}\partial_r a_2 - \frac{1}{r^2}a_2 - \frac{q_0}{r^3}, \\[1mm]  0 = & \ 
        r\left(r^6-m_0r^2+q_0^2\right)\partial_r^2 s_2 - \left(-5r^6+m_0r^2+q_0^2\right)\partial_r s_2 - 2i\omega r^5 \partial_r s_2 - 3i\omega r^4 s_2  \\  & + \frac{1}{3}q^2r^3 s_2 - 36r^4 b_2 - 12 r^5 \partial_r b_2 - \frac{2}{3}r^3 s_1.
\end{aligned}
\end{equation}

\par \medskip
\paragraph{\bf Sector III: $\{b_3,V_3,s_3\}$.}
\begin{equation}
\label{eq:41}
\begin{aligned}
 0{}= & \ 6r^4 \partial_r^2 b_3 + 30r^3 \partial_r b_3 + 6r^4 q_0 \partial_r V_3 + 30r^3 q_0 V_3 - \frac{2}{3} r^2 q^2 \partial_r s_3, \\[1mm] 0= & \
        \left(q_0^2r^2-m_0r^4+r^8\right)\partial_r^2 V_3 + \left(7q_0^2 r-9m_0 r^3+13 r^7\right)\partial_r V_3 + 5\left(q_0^2-3m_0r^2+7r^6\right)V_3 \\ & + 12q_0\partial_r b_3 - 11i\omega r^5 V_3 - 2i\omega r^6\partial_r V_3 + \frac{1}{r}\partial_r a_1 - \frac{1}{r^2}a_1 - \frac{1}{r^2}, \\[1mm] 0 = & \
        r\left(r^6-m_0r^2+q_0^2\right)\partial_r^2 s_3 - \left(-5r^6+m_0r^2+q_0^2\right)\partial_r s_3 - 2i\omega r^5 \partial_r s_3 - 3i\omega r^4 s_3 \\ & + \frac{1}{3}q^2 r^3 s_3 - 36r^4 b_3 - 12 r^5 \partial_r b_3. 
\end{aligned}
\end{equation}

\par \medskip
\paragraph{\bf Sector IV: $\{b_4,V_4,s_4\}$.}
\begin{equation}
\label{eq:42}
\begin{aligned}
0 {} = & \ 6r^4\partial_r^2 b_4 + 30r^3\partial_r b_4 + 6r^4 q_0\partial_r V_4 + 30r^3 q_0 V_4-q^2r^2\partial_rs_4, \\[1mm] 0 = & \
         \left(q_0^2 r^2-m_0r^4+r^8\right)\partial_r^2 V_4 + \left(7q_0^2r-9m_0r^3+13r^7\right)\partial_r V_4 + 5\left(q_0^2-3m_0r^2+7r^6\right)V_4 \\ & + 12q_0\partial_r b_4 - 11i\omega r^5V_4 - 2i\omega r^6\partial_r V_4 - r^4 q^2 V_4 + 8\sqrt{3}q_0 r^2 \kappa V_1+\frac{8\sqrt{3}\kappa q_0^2}{r^5},\\[1mm] 0 = & \
          r\left(r^6-m_0r^2+q_0^2\right)\partial_r^2 s_4 - \left(-5r^6+m_0r^2+q_0^2\right)\partial_r s_4 - 2i\omega r^5 \partial_r s_4 - 3i\omega r^4 s_4-18r^4 b_4-6r^5\partial_rb_4.
\end{aligned}
\end{equation}

\par \medskip
\paragraph{\bf Sector V: $\{a_1,a_2\}$.}
\begin{equation}
\label{eq:43}
\begin{aligned}
0{}= & \ \partial_r a_2 - r\partial_r^2 a_2 + 5r^5 V_1 + r^6\partial_r V_1 - 5q^2r^5 V_2 - r^6q^2 \partial_r V_2, \\[1mm] 0= & \
       \partial_r a_1 - r\partial_r^2 a_1 - 5q^2r^5 V_3 - q^2 r^6 \partial_r V_3.
\end{aligned}
\end{equation}

Once these sectors have been solved, the remaining scalar function $k$ is
obtained by radial integration:
\begin{equation}
\label{eq:44}
\begin{aligned}
\partial_r k
={}&
\Bigg[
2r^2 - \frac{i\omega r}{3}+ q_0 \left(\frac{4a_2}{r^3}-\frac{2\partial_r a_2}{r^2} \right)- 12r^2(b_1-q^2b_2)
\\
&\qquad
- 2r^3 \left( \partial_r b_1  - q^2\partial_r b_2\right)+ \frac{2q^2r}{9} (2s_1-q^2s_2)\Bigg]\partial\beta
\\
&+
\Bigg[q_0 \left( \frac{4a_1}{r^3} - \frac{2\partial_r a_1}{r^2} \right)+ 12r^2q^2b_3+ 2r^3q^2\partial_r b_3- \frac{2q^4r}{9}s_3\Bigg]q^{(0)} .
\end{aligned}
\end{equation}
Equations~\eqref{eq:39}-\eqref{eq:44} constitute the central bulk system used throughout the remainder of the paper. Their solutions, subject to horizon regularity, boundary normalizability, and the Landau-frame condition, determine the normalizable coefficients of the bulk fields. Through Eqs.~\eqref{eq:23} and~\eqref{eq:24}, these coefficients determine the corresponding transport coefficients in the hydrodynamic limit and their frequency- and momentum-dependent all-order generalizations.

\section{Hydrodynamic expansion and analytic results}
\label{section-3}

The all-order radial system derived in Sec.~\ref{section-2} contains the complete linear response at arbitrary frequency and spatial momentum. Before solving this system numerically, it is useful to recover its hydrodynamic limit analytically. We therefore expand the radial coefficient functions in small boundary derivatives and determine the leading transport coefficients. This provides both the low-frequency limit of the generalized transport functions and an analytic check of the all-order construction.

\subsection{Derivative expansion}

To organize the hydrodynamic expansion, we introduce a bookkeeping parameter $\lambda$ through
\begin{equation}
\omega\rightarrow\lambda\omega,
\qquad
q_i\rightarrow\lambda q_i,
\end{equation}
and expand each radial coefficient function in powers of $\lambda$. For example,
\begin{equation}
\label{eq:gradient-b1}
b_1(\rho,\omega,\vec q)
=
\sum_{n=0}^{\infty}
\lambda^n b_1^{(n)}(\rho,\omega,\vec q).
\end{equation}
The same expansion is applied to the remaining coefficient functions,
\begin{equation}
\label{eq:gradient-expansion}
\begin{aligned}
a_i &= \sum_{n=0}^{\infty}\lambda^n a_i^{(n)},
&
b_i &= \sum_{n=0}^{\infty}\lambda^n b_i^{(n)},
&
V_i &= \sum_{n=0}^{\infty}\lambda^n V_i^{(n)},
&
s_i &= \sum_{n=0}^{\infty}\lambda^n s_i^{(n)} .
\end{aligned}
\end{equation}
The parameter $\lambda$ is set to unity after the expansion has been performed.

Because the basis tensors in Eq.~\eqref{eq:37} themselves contain explicit boundary derivatives, the superscript on a radial coefficient does not by itself give the total hydrodynamic order of the corresponding term. For example, $s_1^{(0)}$ multiplies the shear tensor $\sigma_{ij}$ and therefore contributes at first order in derivatives, whereas $s_1^{(1)}$ contributes at second order. We use this convention throughout the perturbative analysis.

At zeroth derivative order the Sector-I equations admit only
\begin{equation}
\label{eq:zero-order}
b_1^{(0)}=0,
\qquad
V_1^{(0)}=0,
\end{equation}
once horizon regularity, boundary normalizability, and the Landau-frame condition are imposed. Thus, as expected, the homogeneous equilibrium background does not generate a dissipative correction. The radial equations leading to Eq.~\eqref{eq:zero-order} are collected in App.~\ref{app:zeroth-order}.

\subsection{First-order transport}

At first order in the hydrodynamic expansion, the relevant structures are the shear tensor $\sigma_{ij}$, the acceleration $\partial_v\beta_i$, the charge gradient $\partial_i q^{(0)}$, the vorticity vector $\ell_i$, and the scalar expansion $\partial_i\beta_i$. The acceleration and charge-gradient contributions arise from the coupled vector sectors and enter
the first-order current through the combination $\partial_i q^{(0)}+3q_0\partial_v\beta_i$. The complete first-order radial systems and their explicit solutions are given in Appendices~\ref{app:first-order-systems} and \ref{app:first-order-bulk-fields}; here we quote only the results required to extract the boundary transport coefficients.

The tensor mode $s_1^{(0)}$ satisfies
\begin{equation}
\label{eq:first-shear-radial}
r^2
\partial_r
\left[
r^5 f(r)\,
\partial_r s_1^{(0)}
\right]
+3r^4
=0 .
\end{equation}
The solution regular at the horizon and normalizable at the boundary is
\begin{equation}
\label{eq:first-shear-integral}
 s_1^{(0)}(\rho)=\frac{1}{R}\int^\infty_\rho \frac{p(p^2+p+1)}{(p+1)(p^4+p^2-M+1)}dp
\end{equation}
in rescaled variables, whose near-boundary behavior is
\begin{equation}
\label{eq:first-shear-asymptotic}
s_1^{(0)}(r)
=
\frac{1}{r}-\frac{R^3}{4r^4}+\mathcal{O}\left(r^{-5}\right) .
\end{equation}
This agrees with the known charged fluid/gravity result ~\cite{Erdmenger:2008rm,Banerjee:2008th}.

The accompanying Sector-I vector modes reduce to
\begin{equation}
\label{eq:first-sector1-vector}
b_1^{(1)}(\omega,q,\rho)=b_1^{(1)}(\omega,q,r)=-\frac{i\omega}{6r}+\frac{i\omega q_0^2(1+M)}{8MRr^6}+\mathcal{O}(r^{-7}),
\quad
V_1^{(1)}(\omega,q,r)=\frac{3i\omega q_0(1+M)}{4MRr^7}+\mathcal{O}(r^{-8}).
\end{equation}
The charge-diffusion sector gives
\begin{equation}
\label{eq:first-diffusion}
b_3^{(0)}
=
-\frac{Q(1+M)}{24MR^4\rho^6}+\mathcal{O}(\rho^{-7}),\qquad
V_3^{(0)}=-\frac{1+M}{4MR^8\rho^7}+\mathcal{O}(\rho^{-8}).
\end{equation}
while the Chern-Simons sector has the exact form
\begin{equation}
\label{eq:first-anomaly}
\begin{aligned}
b_4^{(0)}(r)
&=
\frac{\sqrt{3}\,\kappa q_0^3}{6m_0r^6},
\qquad
V_4^{(0)}(r)
&=
\frac{\sqrt{3}\,\kappa q_0^2}{m_0r^7}.
\end{aligned}
\end{equation}
Finally, the scalar equation for $k$ yields
\begin{equation}
\label{eq:first-k}
k^{(1)}
=
\frac{2}{3}r^3\partial_i\beta_i .
\end{equation}

These solutions contain the complete first-order response required for the boundary stress tensor and current. In particular, the shear mode $s_1^{(0)}$ determines the viscous stress, $V_3^{(0)}$ determines the charge-gradient contribution to the current, and $V_4^{(0)}$ determines the vorticity-induced anomalous current.

Using the holographic prescriptions in Eqs.~\eqref{eq:23} and~\eqref{eq:24}, the stress tensor takes the first-order form
\begin{equation}
\label{eq:first-stress}
T_{\mu\nu}
=
p\left(
\eta_{\mu\nu}
+4u_\mu u_\nu
\right)
-2\eta\,\sigma_{\mu\nu},
\end{equation}
with
\begin{equation}
\label{eq:first-eta}
p=\frac{m_0}{16\pi G_5},\qquad
s=\frac{R^3}{4G_5},\qquad
\eta
=
\frac{R^3}{16\pi G_5}.
\end{equation}
For two-derivative Einstein gravity this corresponds to the standard $\eta/s=1/(4\pi)$ result ~\cite{Policastro:2001yc,Kovtun:2004de}.

The corresponding current may be written as
\begin{equation}
\label{eq:first-current}
J_i
=
n_0\beta_i -D\,\left(\partial_i \delta n+3n_0\partial_v\beta_i\right) +\xi\,\ell_i+\mathcal{O}(\partial^2),
\end{equation}
where
\begin{equation}
\label{eq:first-coefficients}
n=\frac{\sqrt{3}q_0}{16\pi G_5},\qquad \delta n=\frac{\sqrt{3}q^{(0)}}{16\pi G_5}
\qquad
D=\frac{1+M}{4MR},
\qquad
\xi=\frac{3\kappa q_0^2}{16\pi G_5m_0}.
\end{equation}
Equations~\eqref{eq:first-eta} and~\eqref{eq:first-coefficients} reproduce the established first-order transport coefficients of the charged AdS$_5$ black brane~\cite{Erdmenger:2008rm,Banerjee:2008th}. They therefore provide a direct analytic check of our conventions and of the extraction procedure used below for the all-order transport functions.

\subsection{Constraint equations and boundary conservation laws}

The radial dynamical equations determine the transport data, whereas the remaining components of the bulk equations act as constraints. In the present coordinates the Einstein constraints can be written as
\begin{equation}
\label{eq:constraints-Einstein}
\begin{aligned}
g^{rr}E_{vr}+g^{rv}E_{vv}&=0,
\\
g^{rr}E_{rr}+g^{rv}E_{vr}&=0,
\\
g^{rr}E_{ri}+g^{rv}E_{vi}&=0,
\end{aligned}
\end{equation}
and the Maxwell constraint as
\begin{equation}
\label{eq:constraint-Maxwell}
g^{rr}\mathcal{M}_r
+
g^{rv}\mathcal{M}_v
=0.
\end{equation}

At first derivative order these relations reduce to
\begin{equation}
\label{eq:first-constraints}
\begin{aligned}
\partial_v m^{(0)}
&=
-\frac{4}{3}m_0\partial_i\beta_i,
\\
\partial_i m^{(0)}
&=
-4m_0\partial_v\beta_i,
\\
\partial_v q^{(0)}
&=
-q_0\partial_i\beta_i .
\end{aligned}
\end{equation}
They are precisely the linearized conservation equations
\begin{equation}
\label{eq:boundary-conservation}
\partial_\mu T_{(0)}^{\mu\nu}=0,
\qquad
\partial_\mu J_{(0)}^\mu=0.
\end{equation}
Thus the dynamical and constraint sectors play complementary roles: the former determine the constitutive data, while the latter impose the hydrodynamic evolution of the boundary fields.

\subsection{Second-order tensorial correction}

The first correction beyond the leading viscous response arises from the next term in the expansion of the shear-sector tensor mode. Since $s_1^{(1)}$ multiplies $\sigma_{ij}$, it contributes at second order in boundary derivatives.

After inserting the first-order solutions into the Sector-I equations, the second-order system reduces to a single equation for the tensor mode. The full radial system and intermediate integrations are given in App.~\ref{app:second-order-radial-systems}; the resulting equation can be written as
\begin{equation}
\label{eq:second-shear}
\partial_\rho
\left[
\rho^5f(\rho)
\partial_\rho s_1^{(1)}
\right]
=
\frac{i\omega}{R^2}\left[2\rho^3F'(\rho)+3\rho^2F(\rho)-\rho-18(M-1)\partial_\rho\left(\rho^3 B(\rho)\right)\right],
\end{equation}
where
\begin{equation}
\label{eq:F2}
\begin{aligned}
    & {} F(\rho)=\int_\rho^\infty\left(\frac{p(p^2+p+1)}{(p+1)(p^4+p^2-M+1)}\right)dp,\\ &
    B(\rho)=\frac{1}{3}f(\rho)\int_\rho^\infty\frac{dp}{f(p)^2}\left(\frac{1}{p^8}-\frac{3}{4p^7}\left(1+\frac{1}{M}\right)\right)
\end{aligned}
\end{equation}

The details of the radial integration are not needed for the subsequent all-order analysis; only the normalizable coefficient is required. Near the conformal boundary one finds
\begin{equation}
\label{eq:second-shear-asymptot}
s_1^{(1)}(r)
=
-\frac{i\omega R^2}{8r^4}\left(2+\frac{M}{\sqrt{4M-3}}\log\left(\frac{3-\sqrt{4M-3}}
{3+\sqrt{4M-3}}\right)-\frac{3(M-1)}{2M}\right)+\mathcal{O}(r^{-5}).
\end{equation}
The second-order tensorial transport coefficient arising from the diffusion sector can be written as
\begin{equation}
\begin{aligned}
    s_3^{(0)} {} &  =-\frac{12Q}{R^5}\int_\rho^\infty d\zeta\left(\frac{\zeta}{(M-1)-M\zeta^2+\zeta^6}\right) \int_1^\zeta d\xi\left( \partial_\xi\left(\xi^3 B(\xi)\right)\right) \\ &
    = -\frac{\sqrt{M-1}}{8MR^5\rho^4}+\mathcal{O}(\rho^{-7})
    \end{aligned}
\end{equation}

Similarly, the vorticity sector also develops a tensorial transport coupled to the vorticity vector, which can be written as
\begin{equation}
    \pa_\rho\left(\rho^5f(\rho)\pa_\rho s_4^{(0)}\right)=-\frac{3\sqrt{3}\kappa Q^3}{MR^2\rho^4}
\end{equation}
Whose near conformal boundary solution is,
\begin{equation}
\begin{aligned}
    {}  s_4^{(0)}(\rho)& =\frac{\sqrt{3}\kappa Q^3}{MR^2}\int_\rho^\infty\frac{p^2+p+1}{p^2(p+1)(p^4+p^2-M+1)}dp\\ &
    =\frac{\sqrt{3}\kappa (M-1)^{3/2}}{4MR^2\rho^4}+\mathcal{O}(\rho^{-7})
\end{aligned}
\end{equation}
The associated contribution to the boundary stress tensor \eqref{eq:23} is therefore
\begin{equation}
\label{eq:second-stress}
\begin{aligned}
T_{ij}^{\mathrm{diss}(2)}
{} &=\frac{1}{16\pi G_5}\bigg[R^2\left(2+\frac{M}{\sqrt{4M-3}}\log\left(\frac{3-\sqrt{4M-3}}
{3+\sqrt{4M-3}}\right)-\frac{3(M-1)}{2M}\right)\partial_v\sigma_{ij}-\left(\frac{\sqrt{M-1}}{2MR}\right)\Xi_{ij}\\ & +2\sqrt{3}\kappa R^2\frac{(M-1)^{3/2}}{M}\Omega_{ij}\bigg].
\end{aligned}
\end{equation}

Equation~\eqref{eq:second-stress} contains the expected relaxation-type correction to the shear constitutive relation, together with charge-gradient and vortical tensor contributions~\cite{Baier:2007ix,Banerjee:2008th}. 

\subsection{Third-order tensorial correction}
At third derivative order, the tensor structure $\pi_{ij}=\partial_{\langle i}\partial_{j\rangle}(\partial_m\beta_m)$ contributes to the boundary stress tensor through the transport coefficient $\zeta$. While its neutral limit has been determined previously~\cite{Bu:2014ena}, here we calculate its dependence on the background charge and the Chern-Simons coupling in the fully backreacted charged-brane geometry. This extends the neutral result and isolates the anomaly-dependent contribution to this parity-even transport coefficient. We restrict attention to this tensor channel, leaving other third-order corrections outside the present calculation.

The leading coefficient $s_2^{(0)}$ is sourced by the second-order vector profiles $b_2^{(0)}$ and $V_2^{(0)}$. The details of the calculation are presented in App.~\ref{app:third-order}, the resulting equation can be written as,
\begin{equation}
    R^3\pa_\rho\left(\rho^5f(\rho)\pa_\rho s_2^{(0)}(\rho)\right)=12R^2\pa_\rho\left(\rho^3b_2^{(0)}\right)+\frac{2}{3}\rho F(\rho)
\end{equation}
Whose solution using horizon regularity and boundary conditions is,
\begin{equation}
    s_2^{(0)}(\rho)=-\frac{1}{R^3}\int_\rho^\infty\frac{dx}{x^5f(x)}\int_1^x d\xi\left(12R^2\pa_\xi\left(\xi^3b_2^{(0)}(\xi)\right)+\frac{2}{3}\xi F(\xi)\right)
\end{equation}
Whose near conformal boundary solution is
\begin{equation}
\begin{aligned}
s_2^{(0)}(r)={}&-\frac{R}{4r^4}\Bigg\{\frac{2M+3}{12M}-\frac{6\kappa^2(M-1)^2}{M^2}-\frac{\left(1+\sqrt{4M-3}\right)^{3/2}}
{6\sqrt{2}\sqrt{4M-3}}\arctan\!\sqrt{\frac{1+\sqrt{4M-3}}{2}}\\&+\frac{\left(\sqrt{4M-3}-1\right)^{3/2}}{6\sqrt{2}\sqrt{4M-3}}\operatorname{arctanh}\!\sqrt{\frac{\sqrt{4M-3}-1}{2}}-\frac{1}{6\sqrt{4M-3}}\log\!\left(\frac{3+\sqrt{4M-3}}{3-\sqrt{4M-3}}\right)\Bigg\}\\&+\mathcal{O}(r^{-6}).
\end{aligned}
\end{equation}

So the dissipative term corresponding to the $\pi_{ij}$ tensor in the stress tensor at third order can be written as
\begin{equation}\label{eq:third_order_stress}
\begin{aligned}
T_{ij}^{\mathrm{diss}(3)}=\frac{1}{16\pi G_5}\Big[&-R\,\mathcal{Z}(M,\kappa)\,\pi_{ij}+C_{tt}\,\partial_v^2\sigma_{ij}+C_{\nabla}\,\partial^2\sigma_{ij}+C_q\partial_v\Xi_{ij}+C_{\Omega}\,\partial_v\Omega_{ij}+\cdots\Big],
\end{aligned}
\end{equation}
where
\begin{equation}\label{eq:zeta_hydro}
\begin{aligned}
\mathcal{Z}(M,\kappa)={}&\bigg[\frac{2M+3}{12M}-\frac{6\kappa^2(M-1)^2}{M^2}-\frac{\left(1+\sqrt{4M-3}\right)^{3/2}}{6\sqrt{2}\sqrt{4M-3}}\arctan\!\sqrt{\frac{1+\sqrt{4M-3}}{2}}\\
&+\frac{\left(\sqrt{4M-3}-1\right)^{3/2}}{6\sqrt{2}\sqrt{4M-3}}\operatorname{arctanh}\!\sqrt{\frac{\sqrt{4M-3}-1}{2}}-\frac{1}{6\sqrt{4M-3}}\log\!\left(\frac{3+\sqrt{4M-3}}{3-\sqrt{4M-3}}
\right)\bigg].
\end{aligned}
\end{equation}

The explicit $\kappa^2$ term demonstrates that the anomaly also contributes to a parity-even higher-gradient transport coefficient. In Sec.~\ref{section-5}, we show that this contribution is accompanied by a compensating correction to the generalized shear function, leaving the longitudinal combination $2\widetilde{\eta}-q^2\widetilde{\zeta}$ independent of the Chern--Simons coupling at fixed background parameters.

Only the coefficient of $\pi_{ij}$ is determined here. The coefficients $C_{tt}$, $C_{\nabla}$, $C_q$, and $C_{\Omega}$ require higher-order expansions of the remaining transport functions and are left undetermined. The above third-order coefficient for $\pi_{ij}$ matches \cite{Bu:2014ena} when $M=1$.

The analytic results reveal contrasting near-extremal behavior in the two tensor coefficients. Writing the coefficient of $\partial_v\sigma_{ij}$ in Eq.~\eqref{eq:second-stress} as $R^2C_\sigma(M)$, we find 
\begin{equation}
    C_\sigma(M)=\log(3-M)+1-\log9
    +\mathcal{O}\!\left((3-M)\log(3-M)\right).
\end{equation}
Its logarithmic divergence signals that the conventional derivative expansion is not uniformly controlled as extremality is approached. By contrast, $Z(M,\kappa)$ remains finite. Although its inverse-hyperbolic and logarithmic terms diverge individually, their divergent parts cancel, yielding
\begin{equation}
    \lim_{M\to3^-}Z(M,\kappa)
    =\frac14-\frac83\kappa^2
    -\frac{2\sqrt2}{9}\arctan\sqrt2
    +\frac1{18}\log\frac43 .
\end{equation}
For the supersymmetric coupling ($\kappa=1/(2\sqrt{3}$), this limit is approximately $-0.25647$. The near-extremal singularity of the second-order shear coefficient is therefore not shared by the third-order coefficient calculated here. The divergence of $C_\sigma$ signals a loss of uniform control of the conventional derivative expansion near extremality, rather than establishing divergence of the full gradient series. Its sign change alone does not imply a dynamical instability. Similarly, the finite extremal limit of $Z(M,\kappa)$ does not guarantee that subsequent higher-order coefficients in this channel remain finite.

While the perturbative procedure can, in principle, be extended to higher orders, these results illustrate the limitations of a finite-order derivative expansion, particularly near extremality. The increasing complexity of the coupled equations further motivates a numerical approach. We therefore turn in Sec.~\ref{section-4} to a numerical solution of the unexpanded radial equations, retaining their full frequency and momentum dependence. The analytic coefficients derived here provide low-frequency and low-momentum benchmarks for this all-order calculation, including the leading coefficient in the $\pi_{ij}$ channel.

\section{All-order numerical solution}
\label{section-4}

The perturbative analysis of Sec.~\ref{section-3} determines the transport coefficients in the hydrodynamic expansion around $\omega=q=0$. To retain the full frequency- and momentum dependence, we now solve the radial systems \eqref{eq:39}-\eqref{eq:43} numerically without expanding in $\omega$ or $q$. The resulting solutions determine the generalized transport functions discussed in Sec.~\ref{section-5}.

For numerical purposes, we compactify the radial coordinate according to
\begin{equation}
\label{eq:u-coordinate}
u\equiv\frac{1}{\rho},
\qquad
u\in[0,1],
\end{equation}
so that the conformal boundary is located at $u=0$ and the outer horizon at $u=1$. The coupled radial equations are discretized using a Chebyshev pseudospectral collocation method~\cite{boyd01,Dias:2015nua}. We map the interval $u\in[0,1]$ to the standard Chebyshev interval and evaluate the radial functions on a Chebyshev-Gauss-Lobatto grid. Radial derivatives are then represented by the corresponding spectral differentiation matrices, converting each coupled system into a finite-dimensional linear algebraic problem for fixed $(\omega,q,M)$.

The physical solution is fixed by the same conditions used in the analytic construction. At the conformal boundary, non-normalizable modes are excluded so that the boundary metric and external gauge sources remain fixed. At the future horizon, regularity is imposed directly in ingoing Eddington-Finkelstein coordinates, for which the physical
ingoing solution is regular at the horizon. The Landau-frame condition fixes the remaining hydrodynamic ambiguity.

Once the radial system is solved, the generalized transport functions are obtained from the normalizable coefficients in the near-boundary expansions of the corresponding bulk fields. The numerical calculation is therefore the direct all-order counterpart of the analytic extraction performed in Sec.~\ref{section-3}: the hydrodynamic transport coefficients are recovered in the $(\omega,q)\rightarrow(0,0)$ limit, while finite $\omega$ and $q$ incorporate the higher-derivative corrections. The channel-specific normalizations and the relations between the normalizable coefficients and the transport functions are given in Sec.~\ref{section-5}.

The numerical solutions are checked by increasing the spectral resolution and by comparing their small-frequency and small-momentum limits with the analytic results of Sec.~\ref{section-3}. Stable results are obtained with approximately $N\simeq80$~\footnote{We typically use Chebyshev polynomial degree \(N=80\), corresponding to \(N+1=81\) collocation points. Higher resolutions, typically \(N=240\), are used where required near extremality, particularly in the low-frequency region and static momentum sweeps. Convergence is assessed by increasing the radial resolution and comparing the extracted transport functions.} collocation points over the parameter range considered. We study values from the neutral background, $M=1$, up to the near-extremal regime, including $M=2.99$.

The exactly extremal point $M=3$ is not included. At extremality, the blackening factor develops a double zero at the horizon and the near-horizon radial problem changes qualitatively, so the numerical treatment appropriate to a non-extremal horizon cannot simply be continued to $M=3$. The near-extremal solutions considered here remain within the non-extremal problem and allow us to examine how the transport response evolves as this regime is approached.

\section{Transport functions and memory kernels}
\label{section-5}

The numerical solutions of Sec.~\ref{section-4} determine the complete frequency- and momentum-dependent linear response of the charged plasma. In this section, we analyze the corresponding generalized transport functions and their time-domain memory kernels.

The decomposition introduced in Eqs.~\eqref{eq:37}-\eqref{eq:38} provides a useful guide to the physical interpretation of the different channels. The tensor modes $s_1$, $s_2$, $s_3$ and $s_4$ multiply, respectively, the structures
\begin{equation}
\sigma_{ij},
\qquad
\pi_{ij}
=
\partial_{\langle i}\partial_{j\rangle}
(\partial_k\beta_k),
\qquad
\Xi_{ij}
=
\partial_{\langle i}\partial_{j\rangle}q^{(0)},\qquad \Omega_{ij}=\pa_{\langle i}\ell_{j\rangle}.
\end{equation}
while the vector modes $V_3$ and $V_4$ control the charge-gradient and vorticity-induced contributions to the boundary current. We denote the associated generalized transport functions by
\begin{equation}
\widetilde{\eta}(\omega,q^2),\qquad
\widetilde{\zeta}(\omega,q^2),\qquad
\widetilde{\tau}(\omega,q^2),\qquad
\widetilde{D}(\omega,q^2),\qquad
\widetilde{\xi}(\omega,q^2),\qquad \tilde{\lambda}_\Omega(\omega,q^2).
\end{equation}
corresponding, respectively, to the leading shear response, the higher-gradient shear channel, the charge-gradient tensor channel, charge diffusion, and the anomalous vortical current and vortical-gradient tensor channel responses. In the neutral limit, the first two reduce to the two viscosity functions appearing in the all-order linearized constitutive relation of Ref.~\cite{Bu:2014ena}.

Using the holographic prescriptions in Eqs.~\eqref{eq:23} and \eqref{eq:24}, these dimensionless transport functions are extracted directly from the normalizable near-boundary ($u \to 0$) expansion coefficients of the compactified radial fields,
\begin{align}
		\tilde{s}_1(u) &= u + s_1^{(4)} u^4 + \mathcal{O}(u^5), & 
		\tilde{s}_2(u) &= s_2^{(4)} u^4 + \mathcal{O}(u^5), & 
		\tilde{s}_3(u) &= s_3^{(4)} u^4 + \mathcal{O}(u^5), \nonumber \\
        \tilde{s}_4(u) &= s_4^{(4)} u^4 + \mathcal{O}(u^5), &
		\tilde{V}_3(u) &= V_3^{(7)} u^7 + \mathcal{O}(u^8), & 
		\tilde{V}_4(u) &= V_4^{(7)} u^7 + \mathcal{O}(u^8), \label{eq:u_expansions}
\end{align}
through the relations
\begin{equation}
\begin{aligned}
	& {} \tilde{\eta}(\omega, q^2) = -4 s_1^{(4)}, \quad 
	\tilde{\zeta}(\omega, q^2) = -4 s_2^{(4)}, \quad 
	\tilde{\tau}(\omega, q^2) = 4 s_3^{(4)}, \\ & \quad 
	\tilde{D}(\omega, q^2) = - V_3^{(7)}, \quad 
	\tilde{\xi}(\omega, q^2) = \sqrt{3}\, V_4^{(7)},\quad \tilde{\lambda}_\Omega(\omega,q^2)=8s_4^{(4)}. \label{eq:extraction_relations}
	\end{aligned}
	\end{equation}

The transport functions enter physical propagation channels through specific combinations. For a longitudinal velocity perturbation,
$\boldsymbol{\beta}\parallel\mathbf q$, the tensor structures satisfy $\pi_{ij}=-q^2\sigma_{ij}$. Consequently, the velocity-dependent longitudinal viscous stress is governed by the combination
\begin{equation}
    \mathcal{H}_{\mathrm L}(\omega,q^2;\kappa)
    \equiv
    2\widetilde{\eta}(\omega,q^2;\kappa)
    -q^2\widetilde{\zeta}(\omega,q^2;\kappa),
    \label{eq:longitudinal_viscous_function}
\end{equation}
with frequency and momentum expressed in horizon units.

Although both functions individually receive anomaly-dependent corrections, their longitudinal combination satisfies
\begin{equation}
    \mathcal{H}_{\mathrm L}(\omega,q^2;\kappa)
    =
    \mathcal{H}_{\mathrm L}(\omega,q^2;0),
    \label{eq:longitudinal_anomaly_cancellation}
\end{equation}
at fixed background parameters. This follows directly from Eqs.~\eqref{eq:39}--\eqref{eq:40}: the combinations ($b_1-q^2b_2$), ($V_1-q^2V_2$), and ($2s_1-q^2s_2$), together with $a_2$, form a closed radial system in which the explicit Chern--Simons contributions cancel. The same horizon and boundary conditions then imply
Eq.~\eqref{eq:longitudinal_anomaly_cancellation}. This identity holds for the unexpanded radial equations, rather than only at a fixed derivative order.

The cancellation explains why the $\kappa^2$ correction in Eq.~\eqref{eq:zeta_hydro} does not generate an anomaly-dependent longitudinal viscous response. It is consistent with quadratic effective-action analysis in~\cite{Rangamani:2023mok}, which finds anomaly dependence in the transverse sector but not in the sound and charge-diffusion sectors of the quadratic effective action. Our result exhibits this consistency explicitly through the generalized constitutive functions. The conclusion concerns linear response; anomaly-dependent nonlinear couplings to longitudinal modes lie outside the present analysis.

The distinction between these functions in the derivative expansion should not be confused with their temporal behavior. In particular, $\widetilde{\eta}$ and $\widetilde{D}$ provide the leading shear and diffusion responses, whereas $\widetilde{\zeta}$ and $\widetilde{\tau}$ multiply higher-gradient structures. As shown below, the memory kernels of the first pair respond promptly, while those of the higher-gradient channels exhibit a finite buildup time. Remarkably, the anomalous function $\widetilde{\xi}$ also develops a suppressed initial response, even though its leading contribution appears already at first order in derivatives.

Following~\cite{Bu:2015ika}, for any transport function $\widetilde{X}(\omega,q^2)$, we define the corresponding time-domain kernel by
\begin{equation}
\label{eq:memory-kernel}
\widetilde{X}(t,q^2)
=
\frac{1}{\sqrt{2\pi}}
\int_{-\infty}^{\infty}
d\omega\,
\widetilde{X}(\omega,q^2)
e^{-i\omega t}.
\end{equation}
The memory kernel makes the temporal nonlocality of the constitutive relation explicit: the response at a given time depends on the previous history of the hydrodynamic perturbation. Within the numerical accuracy of the Fourier transform, all kernels obtained below vanish for $t<0$.

We vary the background parameter from $M=1$, corresponding to the neutral black brane, to $M=2.99$, close to the extremal value $M=3$. Since $Q^2=M-1$ and the temperature also depends on $M$ through Eq.~\eqref{eq:7}, changing $M$ simultaneously changes the charge density, temperature, and the ratio $\mu/T$. We therefore describe the observed trends in terms of the approach to the near-extremal, large-$\mu/T$ regime rather than attributing them to charge density alone.

\subsection{Leading shear and diffusion channels}

\subsubsection{Generalized shear transport}

The generalized shear transport function $\widetilde{\eta}(\omega,q^2)$ is determined by the normalizable coefficient $s_1^{(4)}$ of the tensor mode $s_1$ through Eq.~\eqref{eq:extraction_relations}. We define the dimensionless shear function by normalizing the shear viscosity with the horizon scale, $\widetilde{\eta}=(16\pi G_5\eta)/R^3$. In the hydrodynamic limit, our result reproduces Eq.~\eqref{eq:first-eta}, while in the neutral limit, it reproduces the all-order result obtained in Ref.~\cite{Bu:2014ena}.

\begin{figure}[t]
    \centering
    \includegraphics[width=\linewidth]{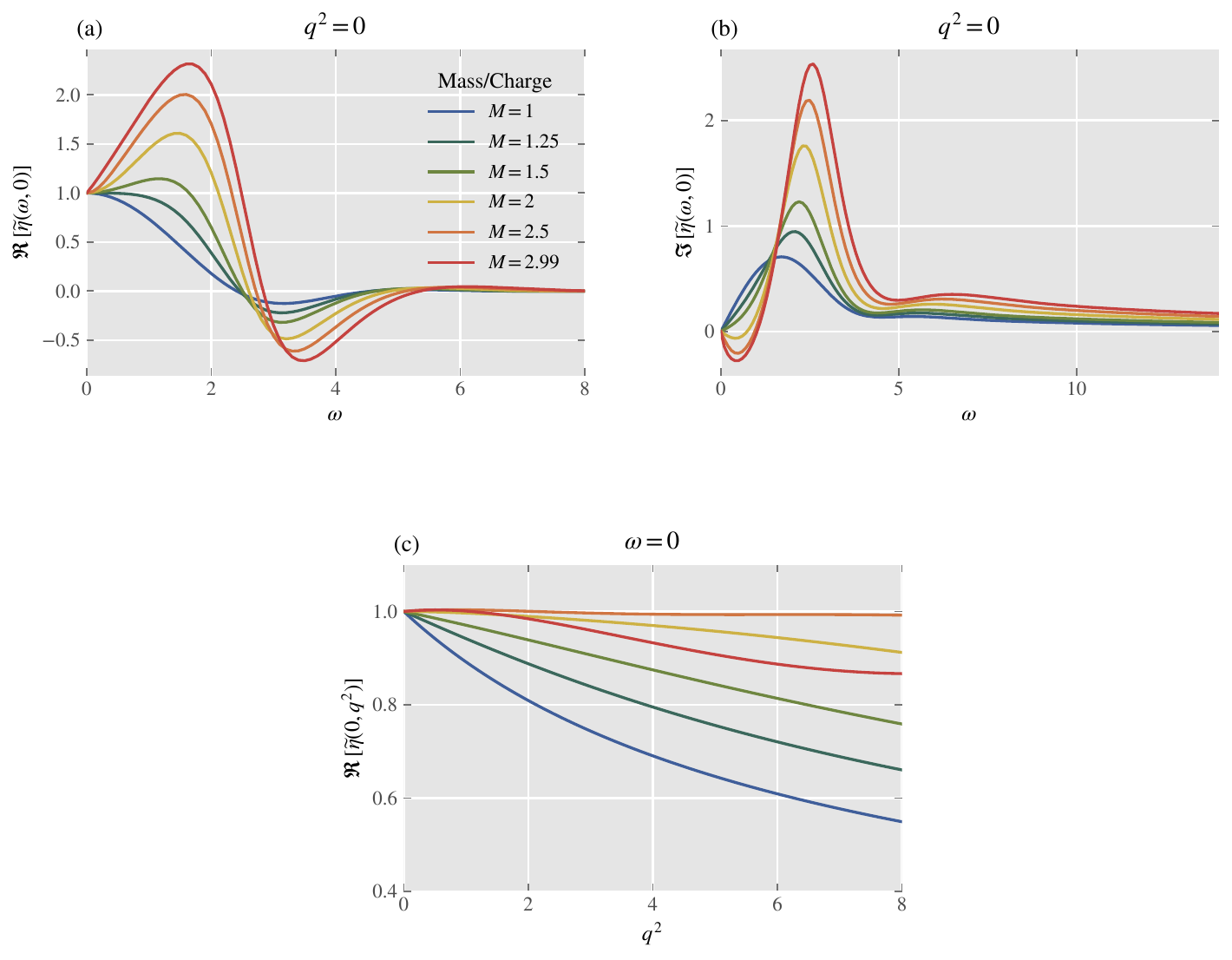}
    \caption{Real and imaginary parts of the generalized shear transport
    function $\widetilde{\eta}(\omega,q^2)$ for different values of $M$.
    The frequency dependence is shown at $q^2=0$, and the momentum
    dependence at $\omega=0$.}
    \label{Fig:1}
\end{figure}

Figure~\ref{Fig:1} shows that all curves approach the same hydrodynamic shear viscosity as $(\omega,q)\rightarrow(0,0)$. Away from this limit, the dependence on the background becomes substantially stronger. In particular, the real part develops a more pronounced intermediate-frequency structure as $M$ approaches extremality, while the imaginary part develops additional low-frequency structure. At $\omega=0$, increasing momentum suppresses the shear response, with the suppression depending increasingly on $M$.

These features indicate that the finite-frequency response cannot be characterized by the hydrodynamic shear viscosity alone. Rather, the full function $\widetilde{\eta}(\omega,q^2)$ is required to describe the dispersive and nonlocal shear response away from the long-wavelength limit.

\begin{figure}[t]
    \centering
    \includegraphics[width=\linewidth]{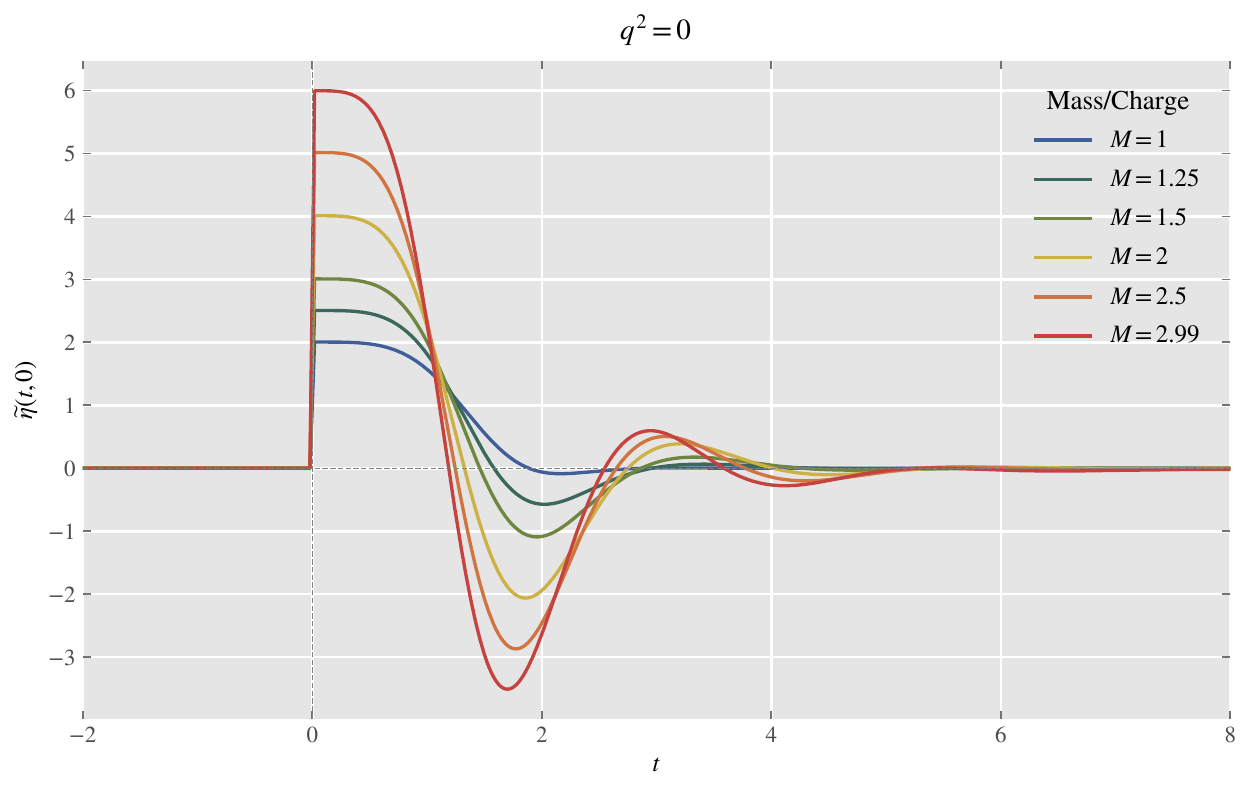}
    \caption{Shear memory kernel $\widetilde{\eta}(t,q^2)$ at $q^2=0$
    for different values of $M$.}
    \label{Fig:2}
\end{figure}

\begin{figure}[t]
    \centering
    \includegraphics[width=\linewidth]{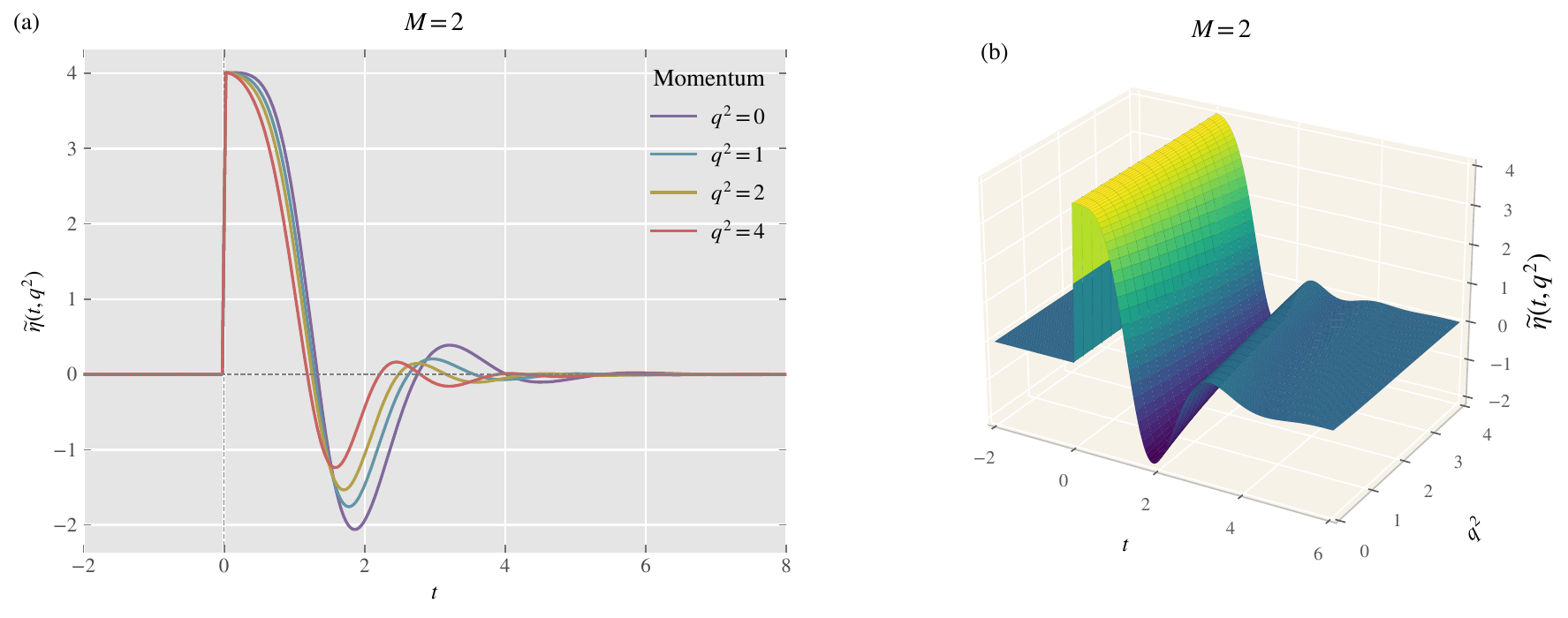}
    \caption{Momentum dependence of the shear memory kernel
    $\widetilde{\eta}(t,q^2)$ at fixed $M=2$.}
    \label{Fig:3}
\end{figure}

The corresponding memory kernels are displayed in Figs.~\ref{Fig:2} and~\ref{Fig:3}. At $q^2=0$, the shear kernel has a finite prompt response followed by a negative overshoot and damped oscillations before decaying to zero. As the near-extremal regime is approached, the amplitude and detailed short-time structure change significantly, while the late-time tail becomes increasingly persistent.

At fixed $M=2$, increasing momentum shortens the memory time, as shown in Fig.~\ref{Fig:3}. The short-time amplitude changes comparatively little, whereas the subsequent relaxation becomes more rapid for larger $q^2$. Thus long-wavelength shear perturbations retain memory for longer than short-wavelength perturbations.

The increasingly persistent late-time behavior toward extremality is qualitatively consistent with the growing importance of the infrared near-horizon dynamics of nearly extremal charged black branes ~\cite{Faulkner:2009wj,Edalati:2009bi,Edalati:2010hk,Davison:2013bxa}. Establishing a direct connection to the AdS$_2$ sector would, however, require a dedicated low-frequency or quasinormal-mode analysis, which we do not attempt here.

\subsubsection{Generalized charge diffusion}

The generalized diffusion function is determined from the normalizable part of the charge-gradient sector governed by Sector III, Eq.~\eqref{eq:41}. We use the dimensionless normalization
\begin{equation}
\widetilde{D}(\omega,q^2)
=
R D(\omega,q^2).
\end{equation}
Its zero-frequency and zero-momentum limit reproduces the analytic coefficient in Eq.~\eqref{eq:first-coefficients}. In the neutral limit, the result agrees with the probe-current calculation of Ref.~\cite{Bu:2015ame}.

\begin{figure}[t]
    \centering
    \includegraphics[width=\linewidth]{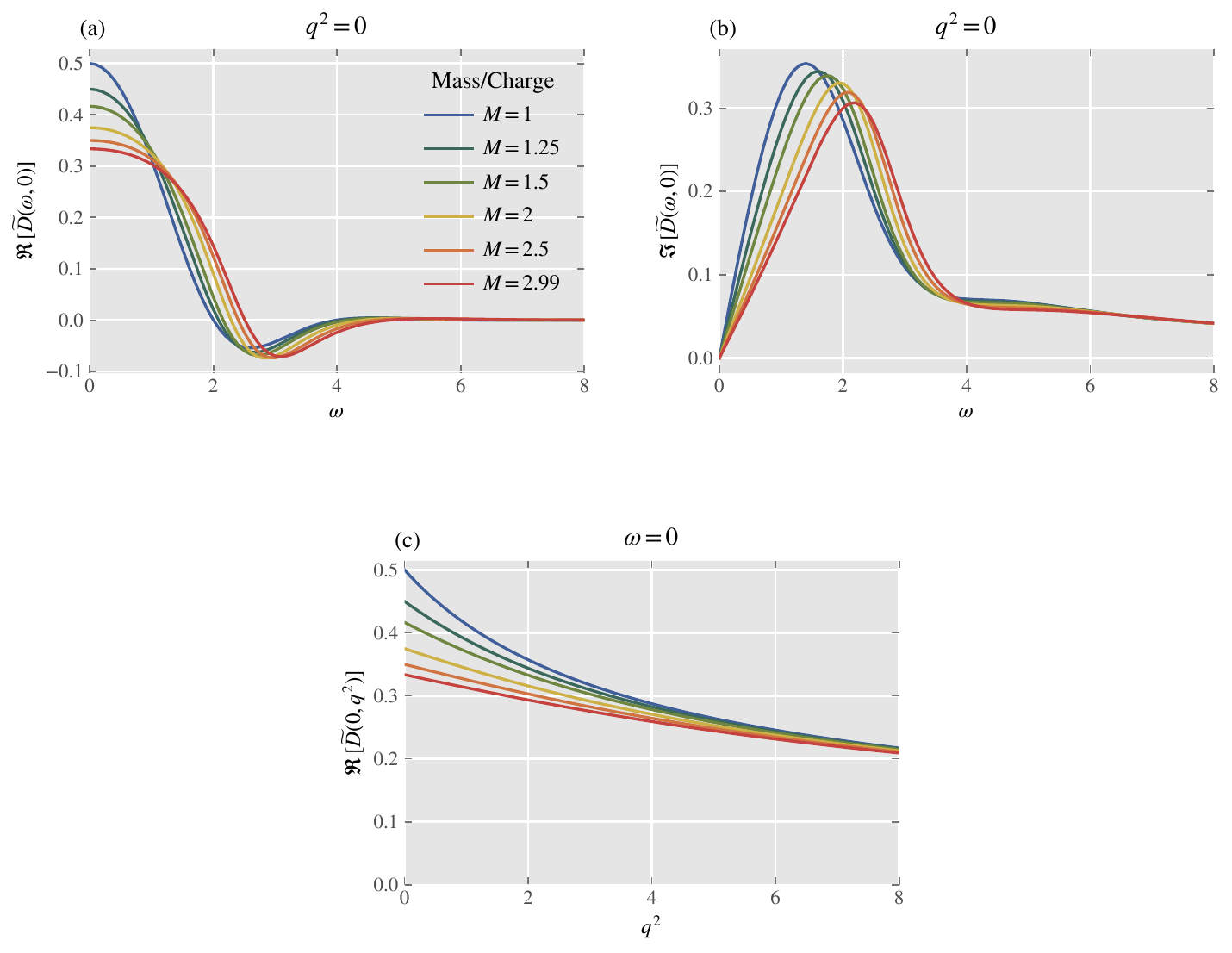}
    \caption{Real and imaginary parts of the generalized charge-diffusion
    function $\widetilde{D}(\omega,q^2)$ for different values of $M$.
    The frequency dependence is shown at $q^2=0$, and the momentum
    dependence at $\omega=0$.}
    \label{Fig:4}
\end{figure}

Unlike the shear viscosity, the hydrodynamic diffusion coefficient itself depends strongly on $M$, in agreement with Eq.~\eqref{eq:first-coefficients}. Figure~\ref{Fig:4} shows that its magnitude decreases as the system approaches the near-extremal regime. Finite frequency produces a shallow negative minimum in the real part, while the overall magnitude of both real and imaginary components is reduced for larger $M$. At fixed $\omega=0$, increasing $q^2$ further suppresses the diffusive response.

\begin{figure}[t]
    \centering
    \includegraphics[width=\linewidth]{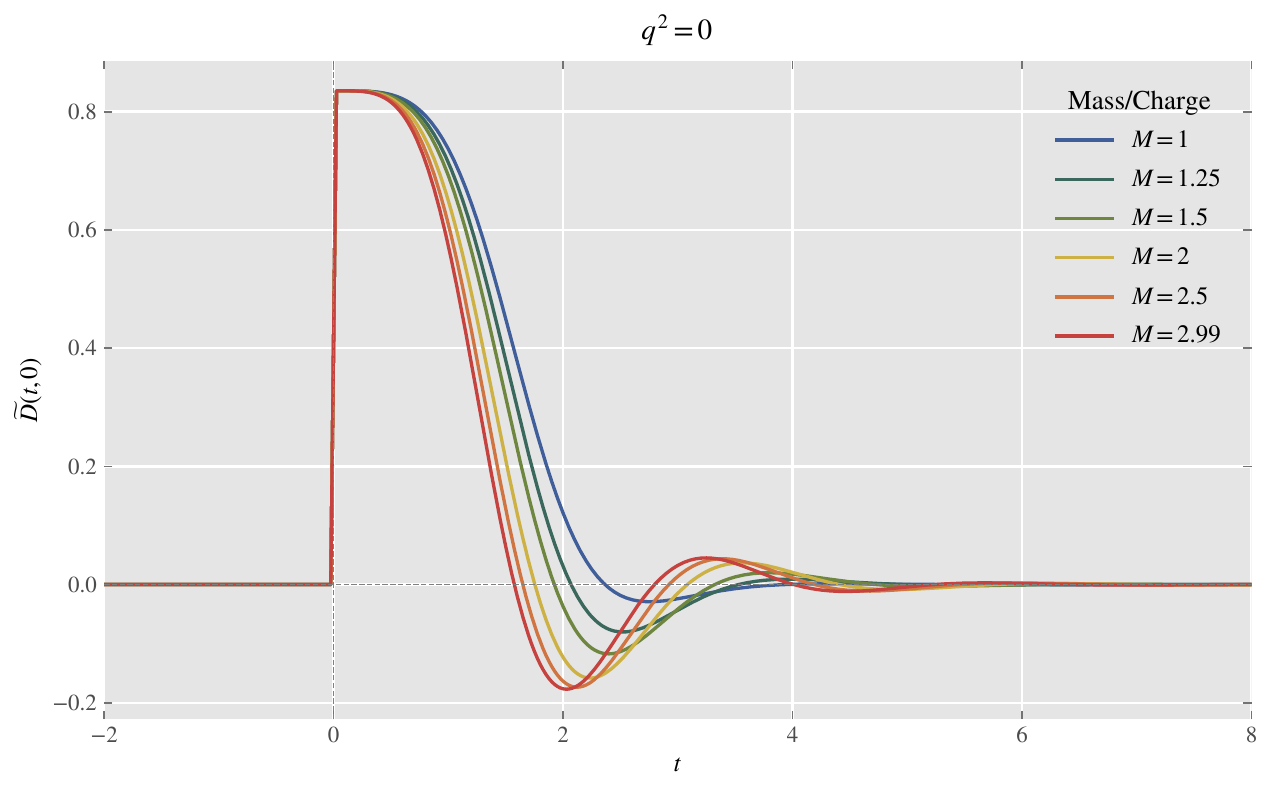}
    \caption{Charge-diffusion memory kernel $\widetilde{D}(t,q^2)$ at
    $q^2=0$ for different values of $M$.}
    \label{Fig:5}
\end{figure}

The time-domain response, shown in Fig.~\ref{Fig:5}, makes the contrast with the shear channel particularly transparent. The initial amplitude of the diffusion remains independent of the parameter $M$, but the dependence on $M$ becomes more apparent at later times showing a pronounced oscillatory response. At the same time, the late-time tail becomes more persistent. Thus a smaller diffusion coefficient does not imply faster loss of memory: transport is suppressed while the associated charge correlations relax more slowly. While the instantaneous diffusive response remains insensitive to $M$, the subsequent temporal evolution develops a more pronounced negative overshoot and a slower decay towards equilibrium accounting for the enhanced memory effects observed near the extremality.
\begin{figure}[t]
    \centering
    \includegraphics[width=\linewidth]{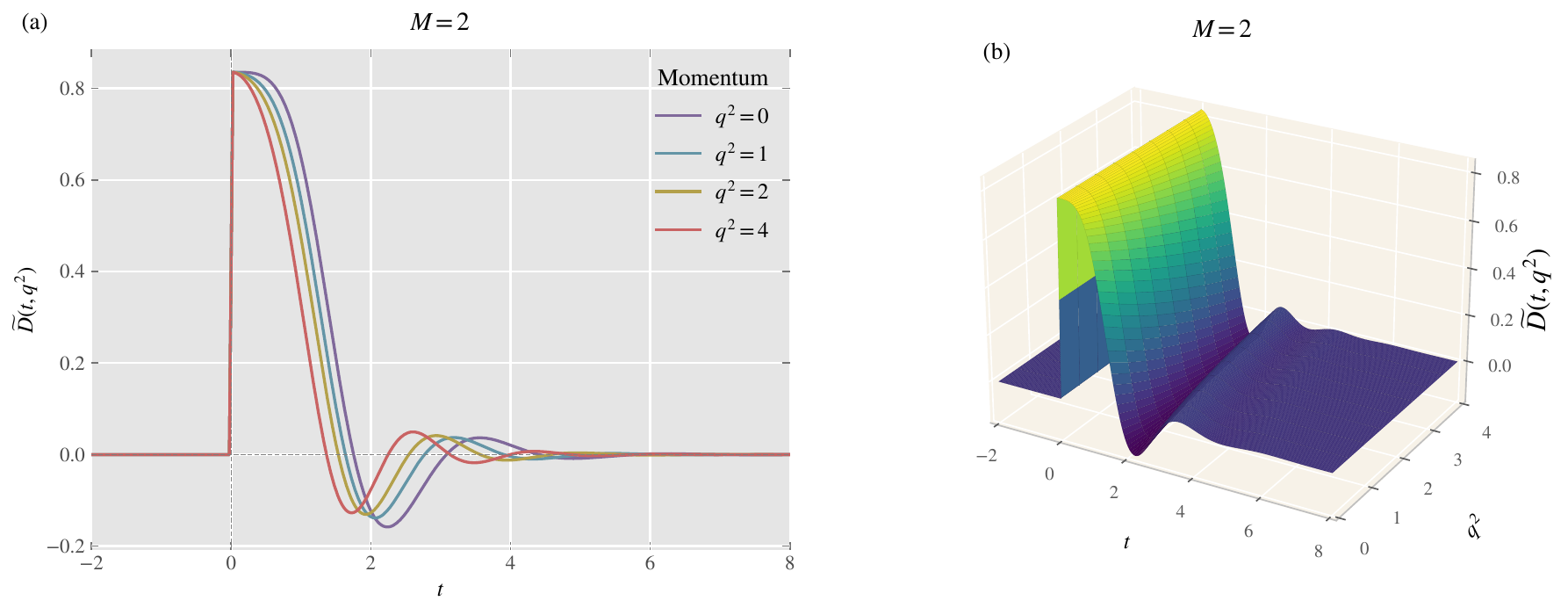}
    \caption{Momentum dependence of the charge-diffusion memory kernel
    $\widetilde{D}(t,q^2)$ at fixed $M=2$.}
    \label{Fig:6}
\end{figure}

Figure~\ref{Fig:6} shows that at small momenta, the memory function exhibits a broader temporal response, with the characteristic minimum occurring at relatively later times. As the momentum is increased, the main response shifts towards earlier times, with the negative overshoot becoming less pronounced. Thus, finite momentum modifies both the characteristic timescale of the initial response and the temporal structure of charge-sector relaxation. 

The leading shear and diffusion channels thus exhibit different near-extremal trends: the shear response develops increasingly pronounced finite-frequency structure, whereas the hydrodynamic diffusion coefficient is suppressed. Nevertheless, both channels develop longer late-time memory as the extremal regime is approached.

\subsection{Higher-gradient transport channels}

The tensor decomposition in Eq.~\eqref{eq:37} contains two additional structures beyond the leading shear term. The function $\widetilde{\zeta}$ multiplies the higher-gradient velocity structure $\pi_{ij}$ and generalizes the second viscosity function introduced in the neutral all-order construction~\cite{Bu:2014ena,Bu:2015ika}. The function $\widetilde{\tau}$ is associated with the charge-gradient tensor $\Xi_{ij}$ and is present only at finite density. Their derivative order is fixed by the corresponding tensor structures.

\subsubsection{Higher-gradient shear function}

The function $\widetilde{\zeta}(\omega,q^2)$ is extracted from the normalizable part of $s_2$, obtained from Sectors I and II. In the neutral limit it reduces to the higher-gradient viscosity function of Ref.~\cite{Bu:2014ena}.

\begin{figure}[t]
    \centering
    \includegraphics[width=\linewidth]{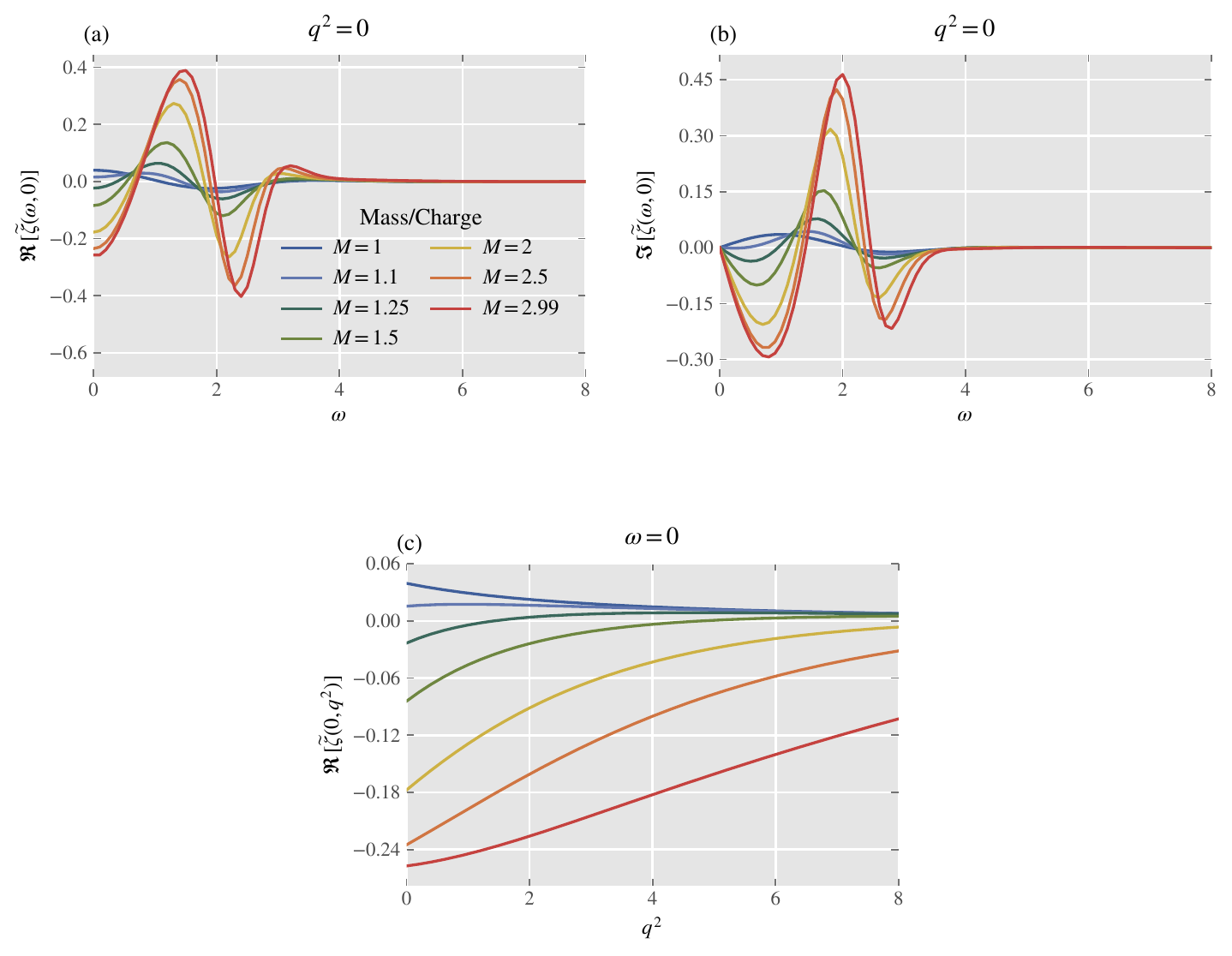}
    \caption{Real and imaginary parts of the higher-gradient shear
    function $\widetilde{\zeta}(\omega,q^2)$ for different values of $M$.}
    \label{Fig:7}
\end{figure}

For $M=1$, Fig.~\ref{Fig:7} reproduces the known neutral result. Finite density strongly modifies both the low-frequency value and the finite-frequency oscillatory structure. In particular, the low-frequency real part changes sign as $M$ is increased, while the amplitudes of the finite-frequency extrema become larger toward extremality. No positivity condition analogous to that of a leading dissipative coefficient is expected for $\widetilde{\zeta}$ because it multiplies a higher-gradient tensor structure.

The magnitude of the coefficient function also decreases rapidly with
increasing momentum. The corresponding stress contribution additionally
depends on the momentum factors carried by $\pi_{ij}$.

\begin{figure}[t]
    \centering
    \includegraphics[width=\linewidth]{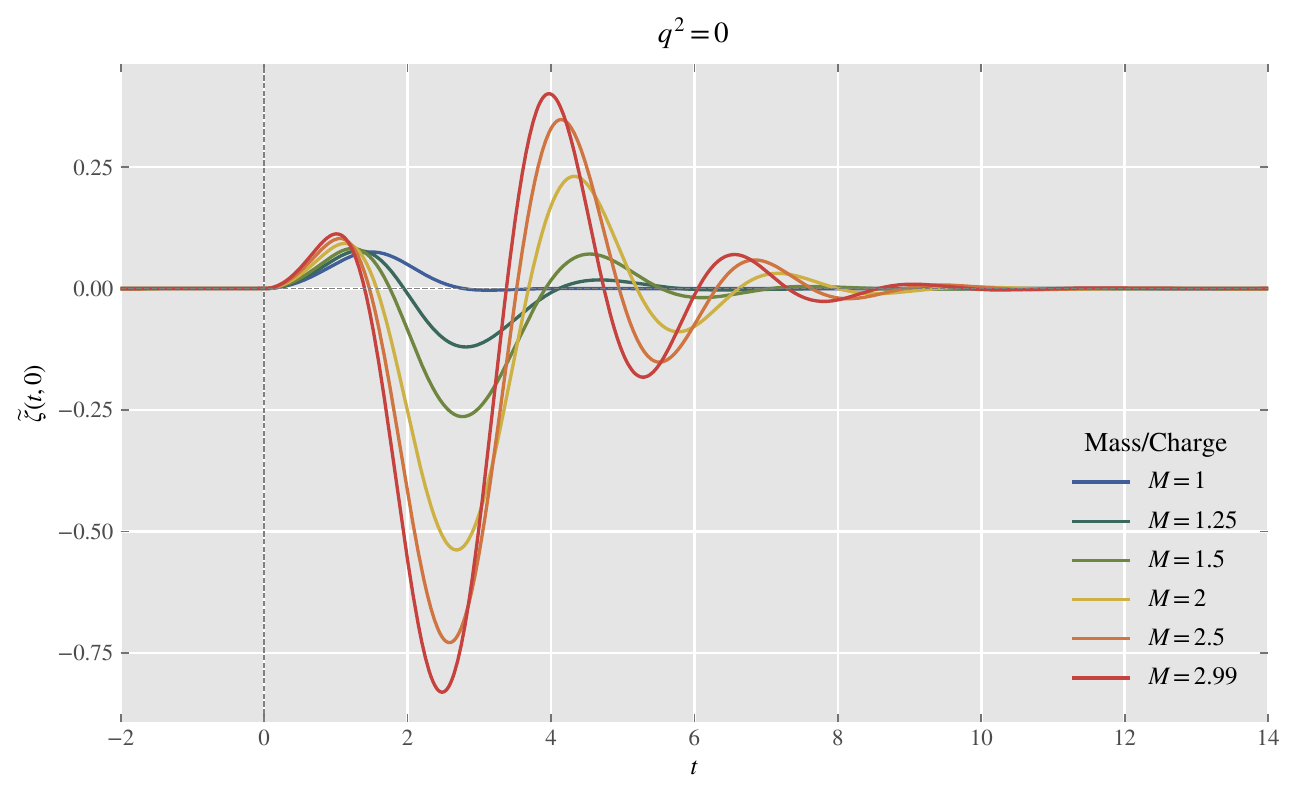}
    \caption{Memory kernel associated with
    $\widetilde{\zeta}(t,q^2)$ at $q^2=0$ for different values of $M$.}
    \label{Fig:8}
\end{figure}

Its time-domain kernel, Fig.~\ref{Fig:8}, differs qualitatively from the leading shear kernel. Rather than displaying a finite prompt response, it is strongly suppressed near $t=0$ and reaches its maximum only after a finite buildup time. The subsequent evolution contains a pronounced negative excursion followed by a positive rebound and damped oscillations, forming a late-time tail before the kernel approaches zero. Increasing $M$ strongly enhances the negative minimum and subsequent oscillatory tail, while the time at which the first extremum occurs changes much less.
\begin{figure}[t]
    \centering
    \includegraphics[width=\linewidth]{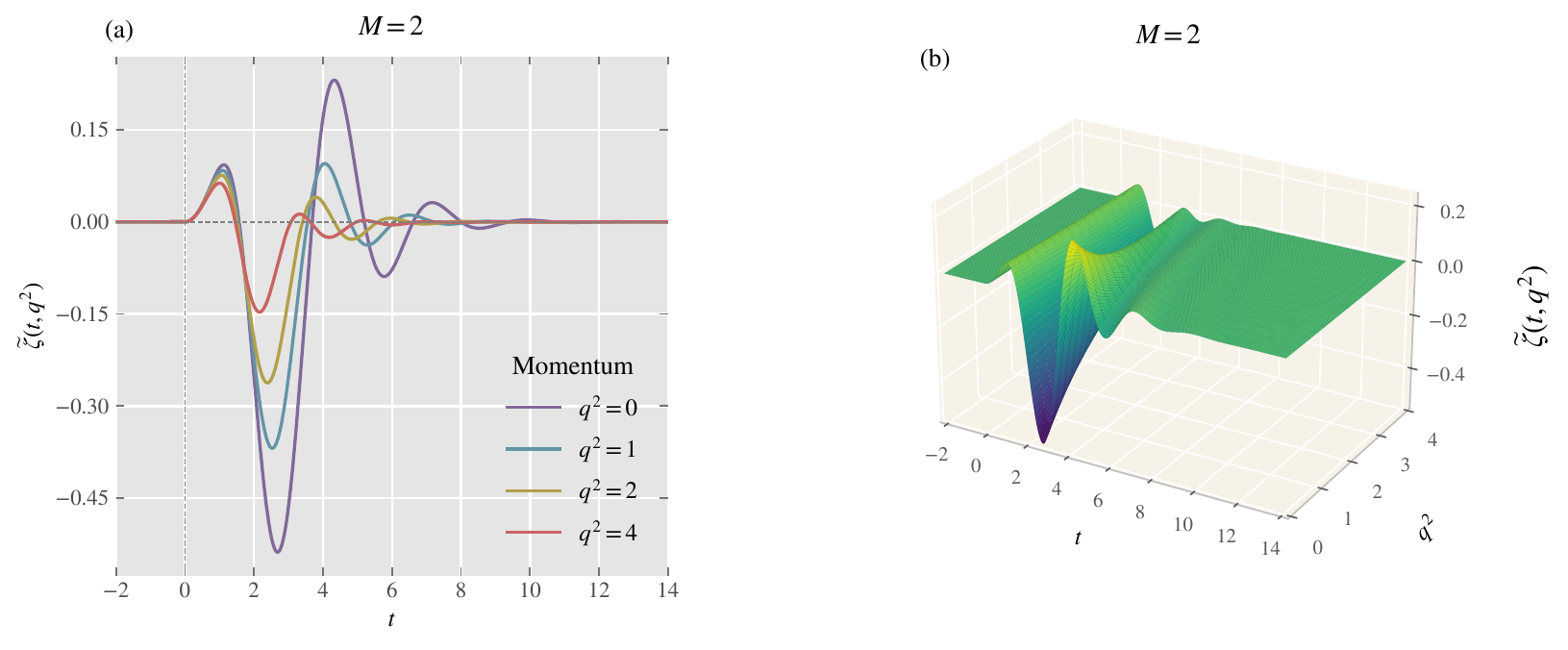}
    \caption{Momentum dependence of the higher-gradient shear memory
    kernel $\widetilde{\zeta}(t,q^2)$ at fixed $M=2$.}
    \label{Fig:9}
\end{figure}

The momentum dependence in Fig.~\ref{Fig:9} follows the same broad pattern as the leading shear kernel: increasing $q^2$ suppresses the amplitude and shortens the late-time tail. The difference between $\widetilde{\eta}$ and $\widetilde{\zeta}$ is therefore primarily in their short-time organization--a prompt response for the leading shear channel and a finite buildup for the higher-gradient correction.

\subsubsection{Charge-gradient tensor function}

The function $\widetilde{\tau}(\omega,q^2)$ is extracted from the normalizable coefficient of $s_3$ in Sector III and multiplies the charge-gradient tensor $\Xi_{ij}$. It therefore belongs to the coupled finite-density charge sector, although it contributes through the stress tensor rather than representing the leading diffusion current itself.

\begin{figure}[t]
    \centering
    \includegraphics[width=\linewidth]{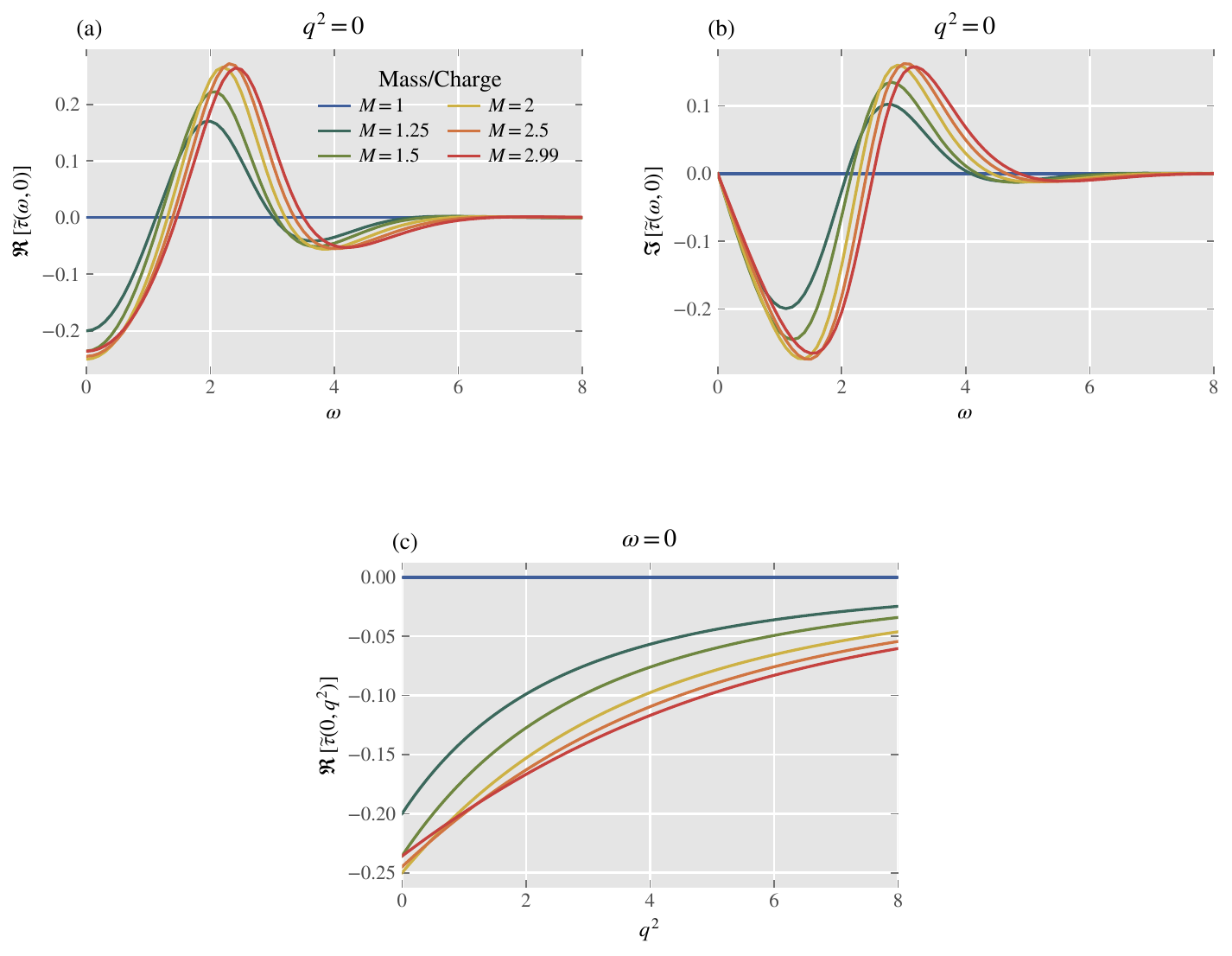}
    \caption{Real and imaginary parts of the charge-gradient tensor
    function $\widetilde{\tau}(\omega,q^2)$ for different values of $M$.}
    \label{Fig:10}
\end{figure}

A useful diagnostic is the neutral limit. As shown in Fig.~\ref{Fig:10}, $\widetilde{\tau}$ vanishes for $M=1$, consistent with the explicit charge dependence of Sector III. The dependence on $M$ differs between the static and finite-frequency responses. The magnitude of the static coefficient, $\widetilde{\tau}(0,0)=-\sqrt{M-1}/(2M)$, increases from zero in the neutral limit, reaches its maximum at $M=2$, and then decreases toward extremality. At finite frequency, however, the response develops more pronounced extrema as $M$ increases. This behavior contrasts with the monotonic suppression of the leading diffusion coefficient $\widetilde D(0,0)$, demonstrating that the leading and higher-gradient charge-sector responses have distinct dependences on the background.

\begin{figure}[t]
    \centering
    \includegraphics[width=\linewidth]{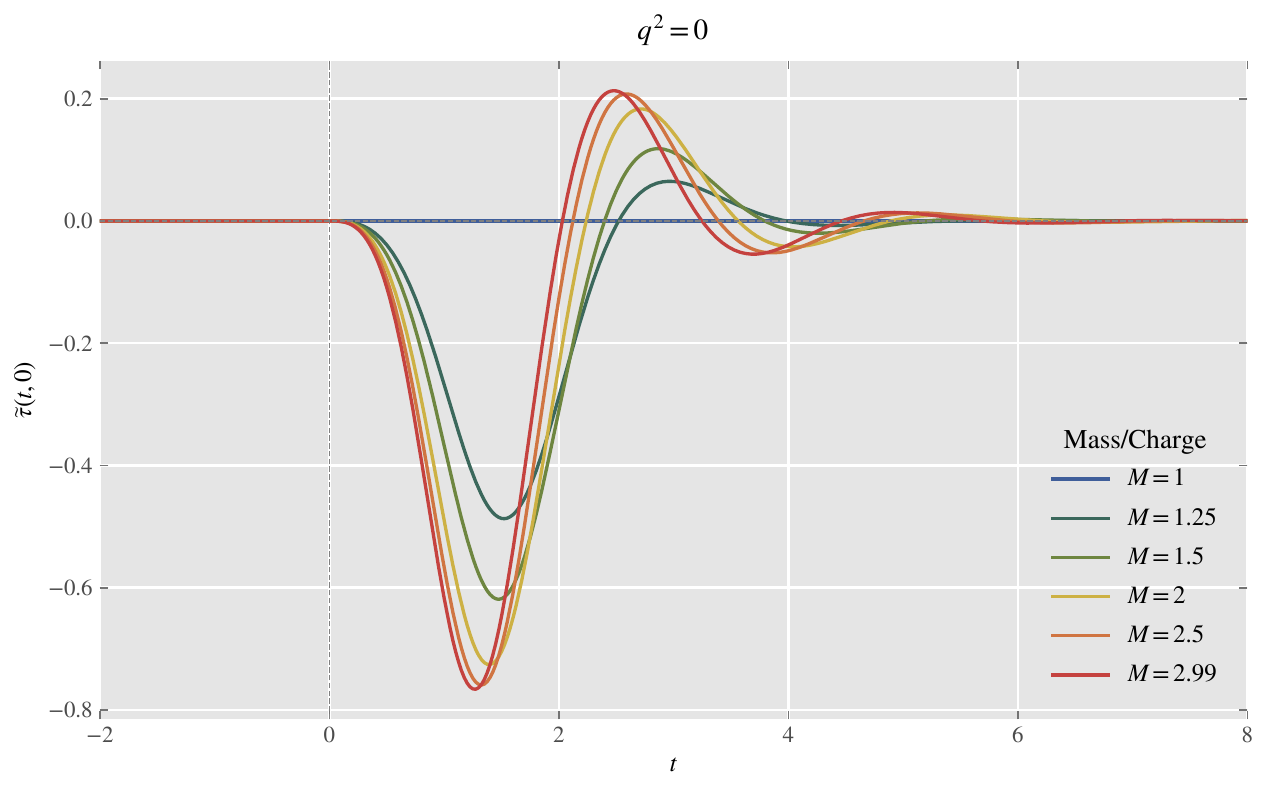}
    \caption{Memory kernel associated with
    $\widetilde{\tau}(t,q^2)$ at $q^2=0$ for different values of $M$.}
    \label{Fig:11}
\end{figure}

The corresponding memory kernel, Fig.~\ref{Fig:11}, is again strongly suppressed at $t=0$ and develops only after a finite buildup time. Its dominant excursion is negative, and its magnitude grows toward the near-extremal regime. At the same time, the late-time relaxation becomes progressively longer lived. The location of the main minimum varies much less strongly than its amplitude.

Comparison with Fig.~\ref{Fig:5} shows that $\widetilde D$ and $\widetilde\tau$ represent different temporal components of the same coupled charge sector: the former controls the prompt diffusive response, whereas the latter encodes a higher-gradient response that develops only after a finite time.

\begin{figure}[t]
    \centering
    \includegraphics[width=\linewidth]{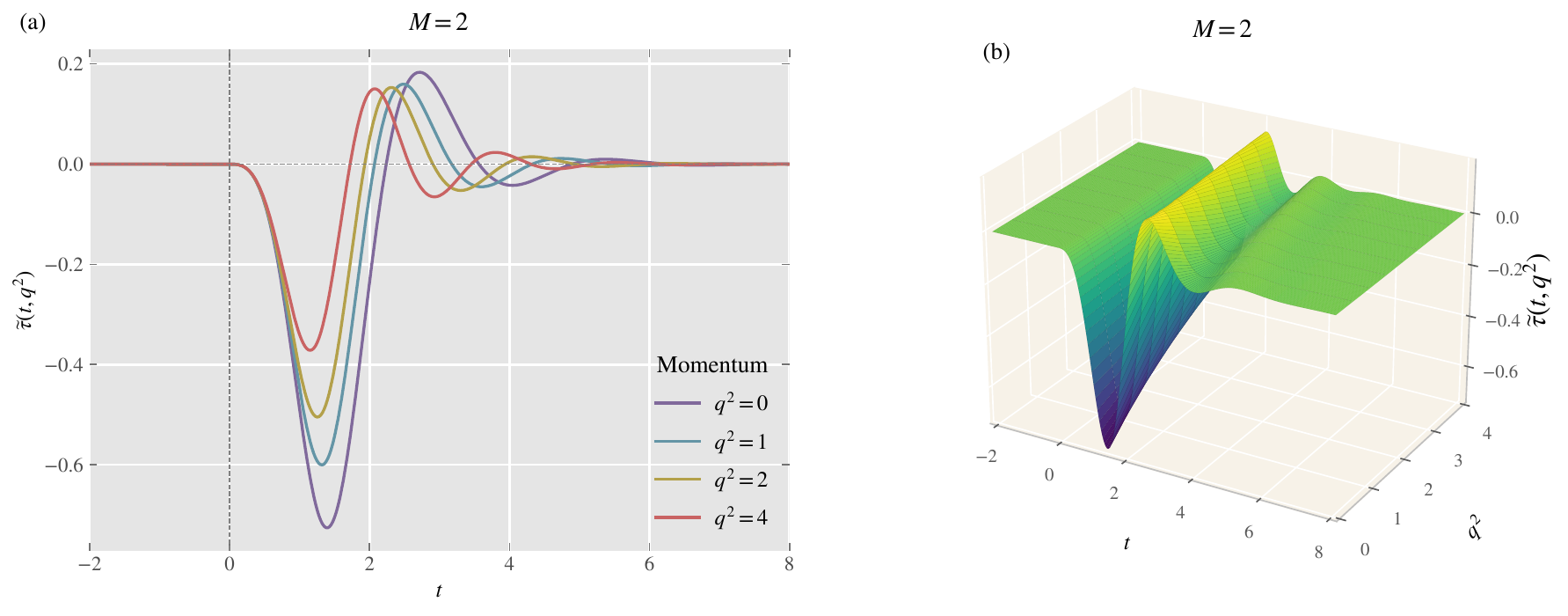}
    \caption{Momentum dependence of the
    $\widetilde{\tau}(t,q^2)$ memory kernel at fixed $M=2$.}
    \label{Fig:12}
\end{figure}

As shown in Fig.~\ref{Fig:12}, increasing momentum reduces both the magnitude and temporal persistence of this suppressed initial response. The higher-gradient charge channel is therefore also predominantly a long-wavelength effect.

\subsection{Anomalous chiral-vortical transport}
\subsubsection{Vortical current function}
The Chern-Simons interaction introduces an anomalous vortical current whose leading hydrodynamic coefficient was recovered analytically in Sec.~\ref{section-3}. The equilibrium chiral-vortical effect is well established in anomalous hydrodynamics and holography~\cite{Erdmenger:2008rm,Banerjee:2008th,Son:2009tf,Jensen:2012jy,Landsteiner:2016led}. Frequency- and momentum-dependent anomalous magnetic transport has also been studied holographically in Refs.~\cite{Bu:2016oba,Bu:2016vum}, while the out-of-equilibrium chiral-vortical effect and frequency dependence of vortical conductivity was investigated in Ref.~\cite{Morales-Tejera:2020xuv,Landsteiner:2013aba}. Here we determine the corresponding full frequency- and momentum-dependent response within the present all-order linearized fluid/gravity framework.

The generalized vortical function $\widetilde{\xi}(\omega,q^2)$ is obtained from the normalizable part of the $V_4$ solution in Sector IV. Its hydrodynamic limit reproduces the coefficient $\xi$ in Eq.~\eqref{eq:first-coefficients}.

\begin{figure}[t]
    \centering
    \includegraphics[width=\linewidth]{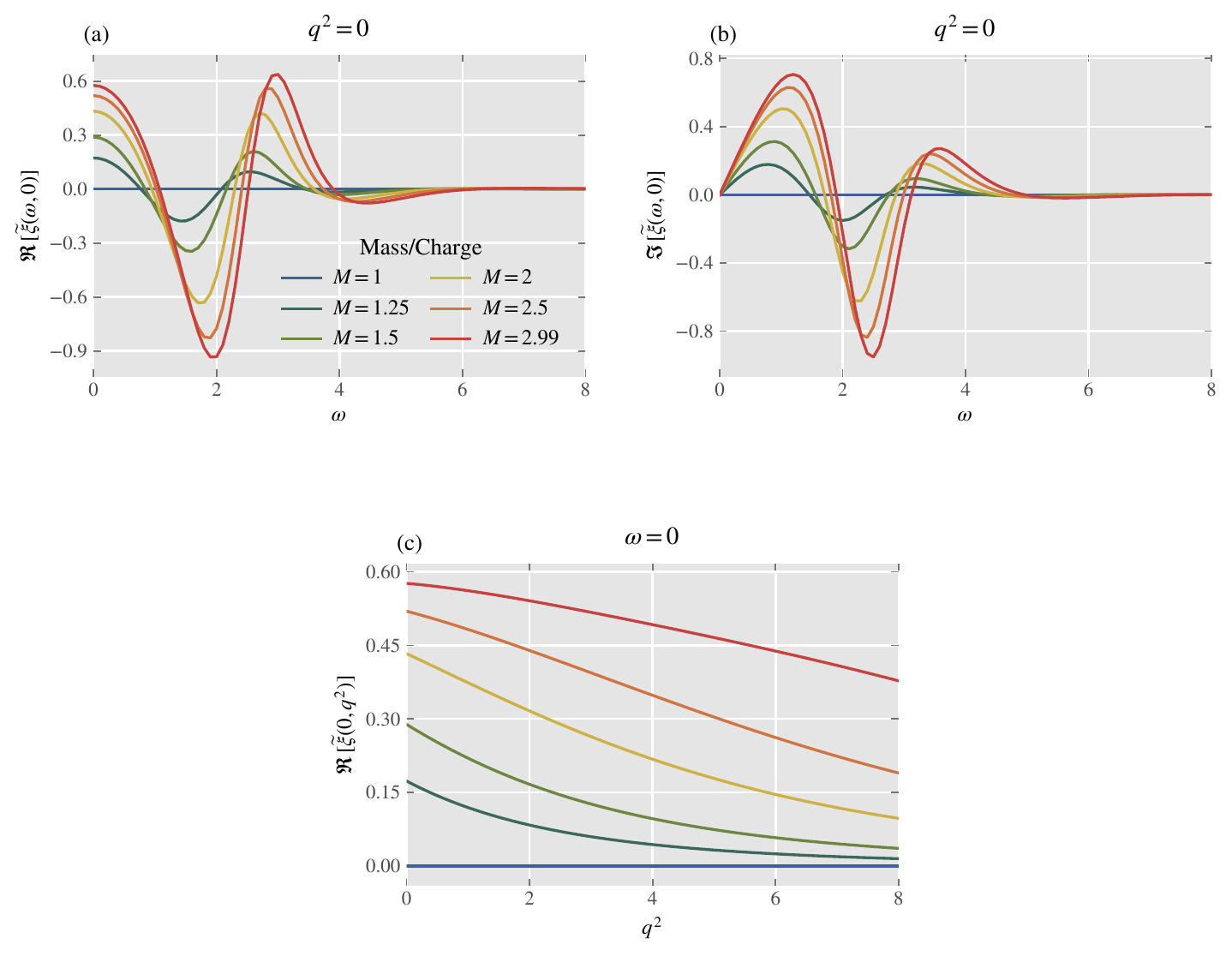}
    \caption{Frequency and momentum dependence of the generalized
    chiral-vortical transport function $\widetilde{\xi}(\omega,q^2)$
    for different values of $M$.}
    \label{Fig:13}
\end{figure}

The anomalous response vanishes in the neutral background and grows as the system is driven toward the charged regime. Figure~\ref{Fig:13} shows a non-monotonic frequency dependence
of the generalized vortical transport function. At $q^2=0$, the real part decreases from its positive static value, crosses zero, and develops a pronounced negative minimum,
followed by a positive maximum and a smaller negative undershoot before approaching zero at larger frequencies. The imaginary part similarly exhibits a positive maximum,
a negative minimum, and a secondary positive maximum. The amplitudes of these features increase substantially with $M$, while the principal extrema shift towards higher
frequencies. The response vanishes in the neutral background, $M=1$, consistently with its anomaly-induced origin in the charged plasma. At $\omega=0$, the real part decreases
monotonically with $q^2$, indicating suppression of the static vortical response at shorter wavelengths. This suppression is less pronounced near extremality, where an appreciable response persists over the momentum range shown.

\begin{figure}[t]
    \centering
    \includegraphics[width=\linewidth]{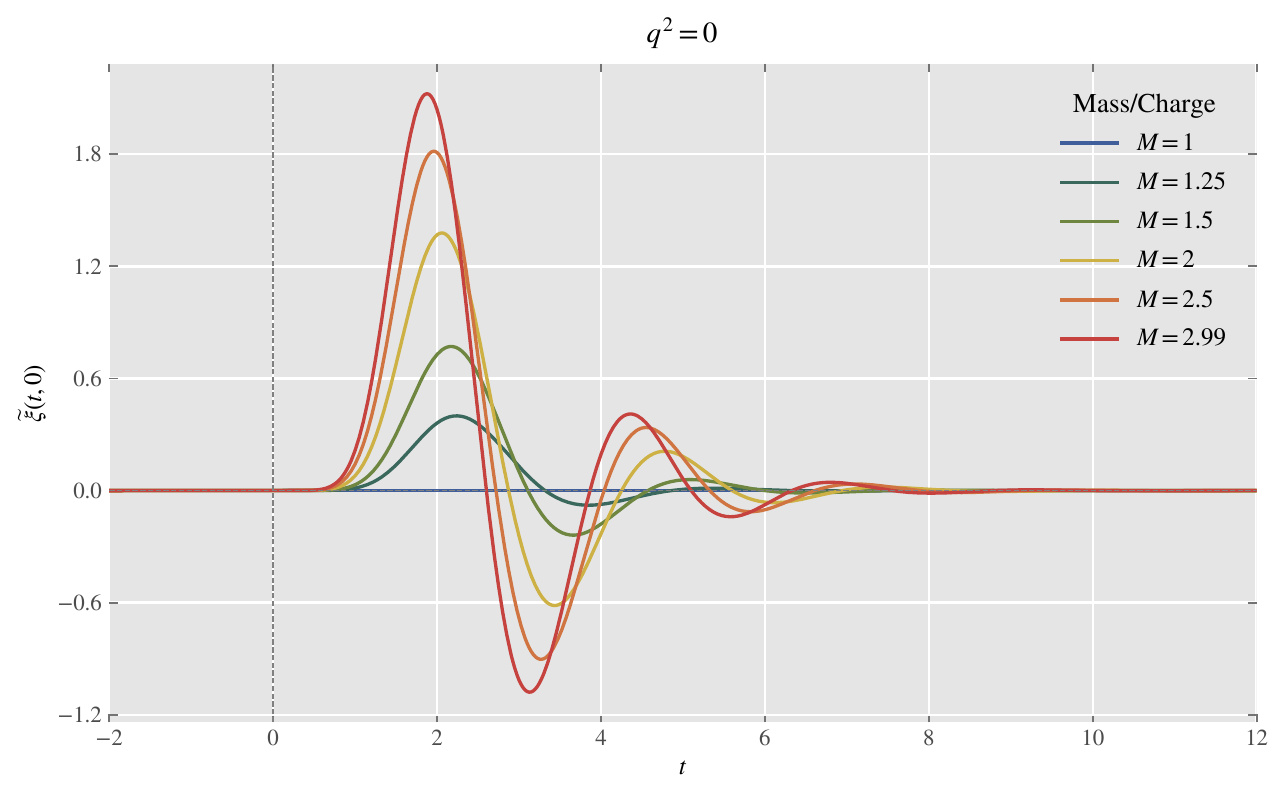}
    \caption{Chiral-vortical memory kernel
    $\widetilde{\xi}(t,q^2)$ at $q^2=0$ for different values of $M$.}
    \label{Fig:14}
\end{figure}

The memory kernel in Fig.~\ref{Fig:14} reveals an important distinction between derivative order and temporal onset. Although the chiral-vortical term is already present at first order in the hydrodynamic expansion, its time-domain response does not show the prompt behavior of $\widetilde{\eta}$ or $\widetilde D$. Instead, its kernel rises gradually from a
near-zero initial value and reaches a positive maximum at finite time. This buildup is followed by a pronounced negative minimum and subsequent oscillations of decreasing amplitude,
before the kernel approaches zero.

Increasing $M$ enhances both the initial positive peak and the negative undershoot, making the oscillatory transient more pronounced. The principal extrema also shift modestly towards
earlier times. The neutral background, $M=1$, has a vanishing response. Thus, the anomalous channel exhibits a finite-time buildup followed by damped, sign-changing relaxation.

\begin{figure}[t]
    \centering
    \includegraphics[width=\linewidth]{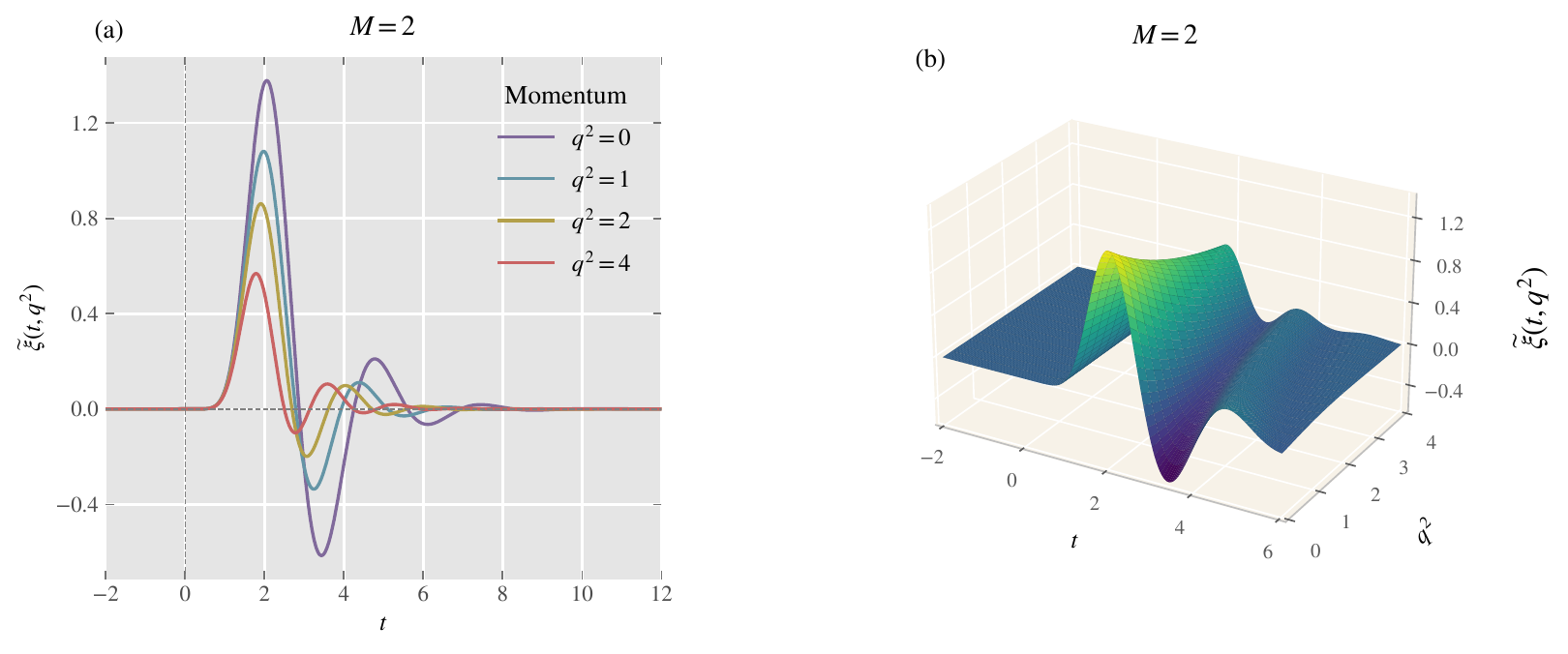}
    \caption{Momentum dependence of the chiral-vortical memory kernel
    $\widetilde{\xi}(t,q^2)$ at fixed $M=2$.}
    \label{Fig:15}
\end{figure}

At fixed $M=2$, Fig.~\ref{Fig:15} shows that increasing momentum suppresses both the initial positive maximum and the subsequent negative undershoot. The principal extrema shift towards earlier times, while the later oscillations become smaller and the kernel approaches
zero sooner over the range shown. Thus, the vortical memory is strongest and most persistent at long wavelengths, although a finite response remains at nonzero momentum. This momentum suppression is consistent with the static Fourier-space behavior in Fig.~\ref{Fig:13}.

\subsubsection{Vortical tensor function}

We finally consider the tensor response associated with the vorticity-gradient structure
\begin{equation}
\Omega_{ij}
=\frac{1}{2}\left(\partial_i\ell_j+\partial_j\ell_i\right),
\qquad
\ell_i=\epsilon_{ijk}\partial_j\beta_k .
\end{equation}
This contribution is governed by Sector IV, Eq.~\eqref{eq:42}, which contains the coupled radial functions $\{b_4,V_4,s_4\}$. The tensor mode $s_4$ enters the metric decomposition through $ \alpha_{ij}\supset2s_4\Omega_{ij}$, and its normalizable boundary coefficient determines the generalized transport function $\widetilde{\lambda}_{\Omega}(\omega,q^2)$.

The tensor equation is sourced by the metric vector mode $b_4$, which is coupled to the gauge-field mode $V_4$ responsible for the vortical current. The Chern--Simons interaction also couples this subsystem to Sector I through the vector modes $V_1$ and $V_4$. Consequently,
$\widetilde{\lambda}_{\Omega}$ is obtained from the coupled Sector-I and Sector-IV system rather than from an independent tensor equation. Since $\Omega_{ij}$ contains two boundary derivatives, its leading contribution appears at second hydrodynamic order. Retaining the full frequency and momentum dependence resums the higher-derivative
corrections to this anomaly-induced tensor response.

\begin{figure}[t]
    \centering
    \includegraphics[width=\linewidth]{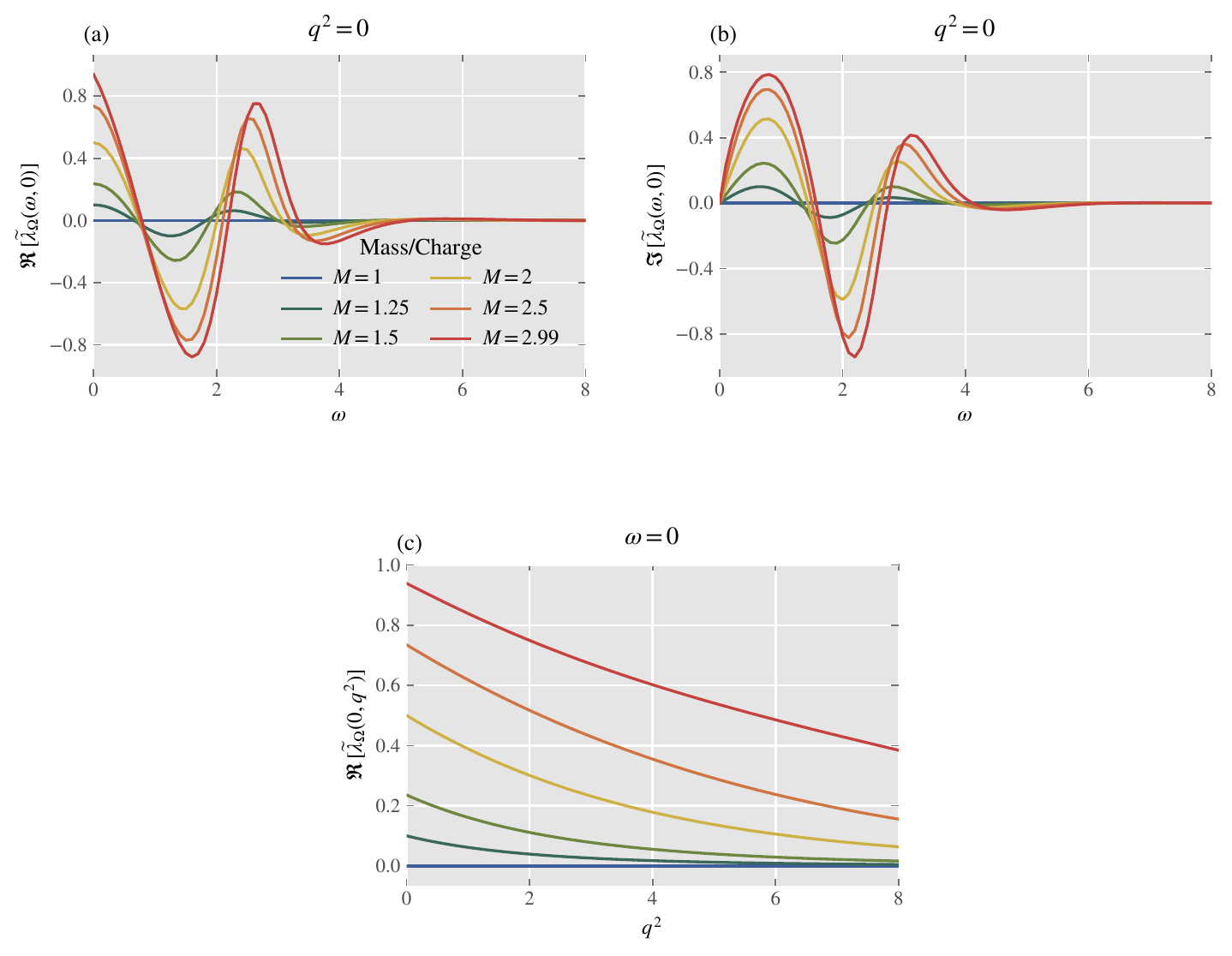}
    \caption{Frequency and momentum dependence of the generalized
    chiral-vortical-gradient transport function $\widetilde{\lambda}_\Omega(\omega,q^2)$
    for different values of $M$.}
    \label{Fig:16}
\end{figure}

The Fourier spectra of $\widetilde{\lambda}_{\Omega}$ in Fig.~\ref{Fig:16} show multiple sign changes and finite-frequency extrema in both their real and imaginary parts. Increasing $M$ enhances their amplitudes, while the response vanishes at $M=1$. These features resemble those of
the chiral vortical current, consistent with the coupling between the vector and tensor modes,
although their relative amplitudes differ. At $\omega=0$, the tensor response decreases monotonically with momentum, with weaker relative suppression near extremality.

\begin{figure}[t]
    \centering
    \includegraphics[width=\linewidth]{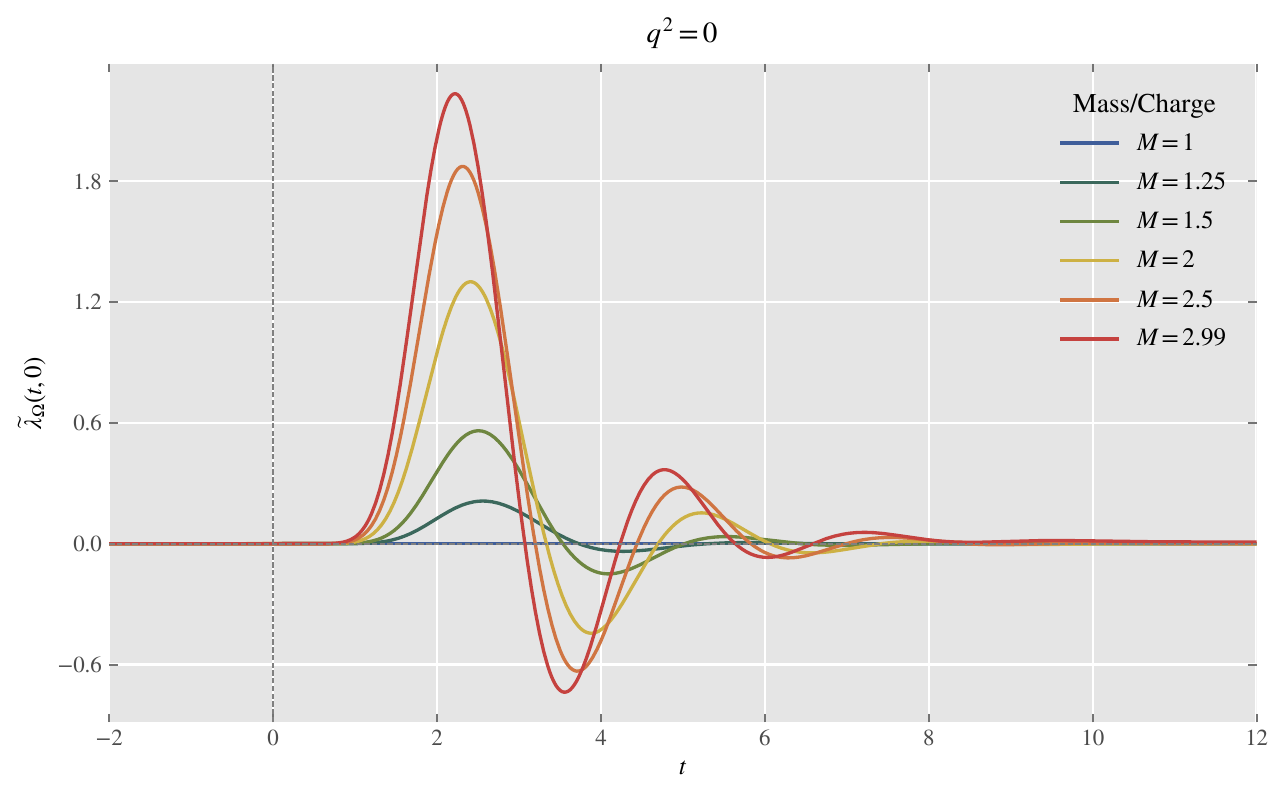}
    \caption{Vortical-gradient tensor memory kernel
    $\widetilde{\lambda}_\Omega(t,q^2)$ at $q^2=0$ for different values of $M$.}
    \label{Fig:17}
\end{figure}
The memory kernel of $\widetilde{\lambda}_{\Omega}$ shown in Fig.~\ref{Fig:17} exhibits a finite-time buildup followed by damped, sign-changing oscillations, resembling the chiral vortical current kernel $\widetilde{\xi}$. Increasing $M$ enhances the positive peak and subsequent oscillations, while the response vanishes at $M=1$. Compared with $\widetilde{\xi}$, the tensor
kernel reaches its principal extrema at later times and has a weaker negative undershoot relative to its initial positive maximum. These differences show that the coupled vector and tensor responses have distinct temporal profiles, rather than differing solely by an overall normalization.

\begin{figure}[t]
    \centering
    \includegraphics[width=\linewidth]{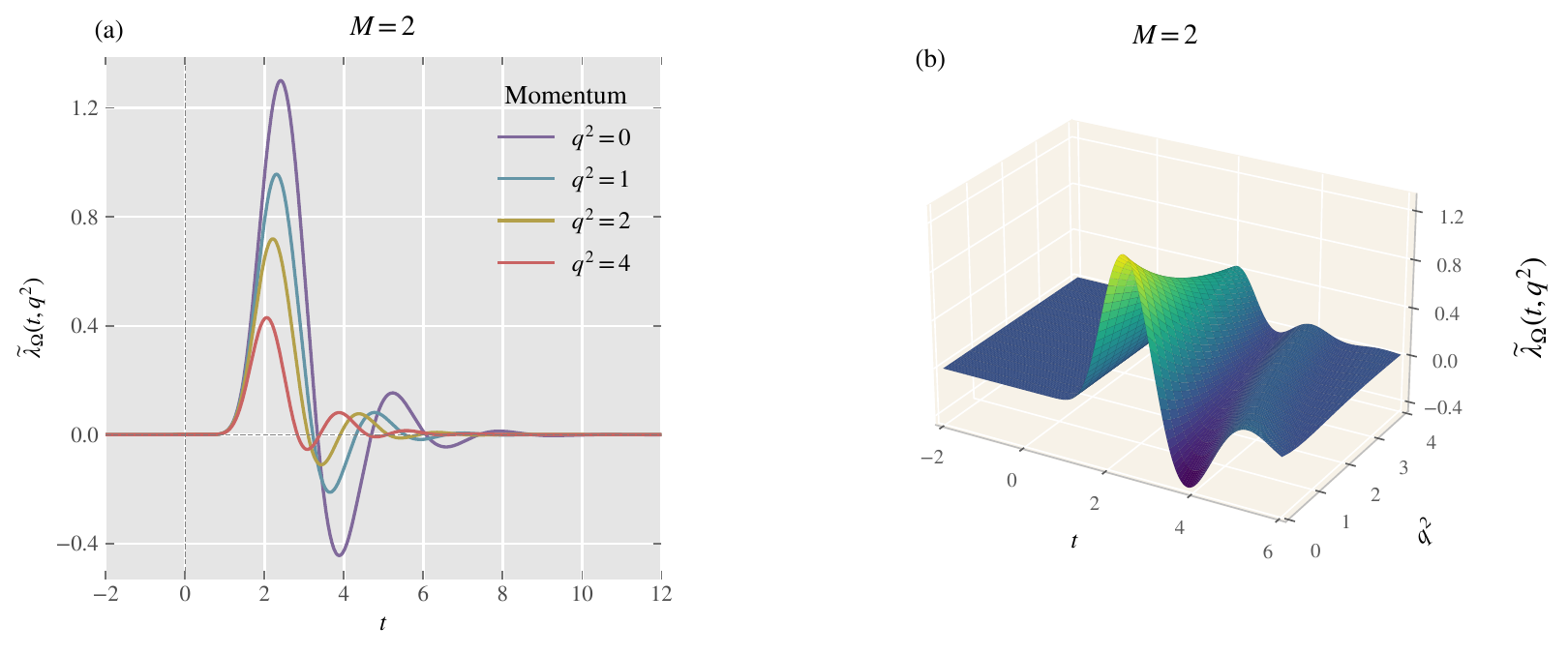}
    \caption{Momentum dependence of the Vortical-gradient tensor memory kernel
    $\widetilde{\lambda}_\Omega(t,q^2)$ at fixed $M=2$.}
    \label{Fig:18}
\end{figure}

At fixed $M=2$, increasing momentum suppresses the positive maximum and negative undershoot of $\widetilde{\lambda}_{\Omega}$, while shifting the principal extrema towards earlier times. The subsequent oscillations also become smaller. These trends resemble those of the chiral vortical current kernel $\widetilde{\xi}$, but the tensor response peaks later and
exhibits a weaker relative negative undershoot. Thus, both channels show their strongest oscillatory memory at long wavelengths, with distinct temporal profiles reflecting their coupled dynamics.
\subsection{Comparison of the transport channels}

The time-domain analysis exposes a hierarchy that is less apparent from the frequency-dependent transport functions alone. The leading shear and diffusion channels, $\widetilde{\eta}$ and $\widetilde D$, exhibit a prompt response. The higher-gradient functions $\widetilde{\zeta}$ and $\widetilde{\tau}$ instead develop after a finite buildup time. The anomalous vortical current $\widetilde{\xi}$ and the vorticity-gradient tensor response $\widetilde{\lambda}_{\Omega}$ also display finite-time buildup, followed by damped, sign-changing oscillations. Their similar qualitative behavior reflects their coupled bulk dynamics, while differences in peak times and relative undershoots distinguish the two responses. Thus, derivative order alone does not determine the temporal profile of a transport channel.

A second common feature is the enhancement of late-time memory as the near-extremal regime is approached. This trend is visible in the shear, diffusion, higher-gradient, and anomalous sectors, although their short-time amplitudes behave differently. The observation is consistent with increasingly slow infrared dynamics near extremality, but a direct identification of the underlying modes would require an analysis of the complex-frequency spectrum.

Finally, increasing spatial momentum generally suppresses the magnitude and duration of the memory kernels. The longest-lived responses are therefore associated with long-wavelength perturbations, whereas short-wavelength fluctuations relax more rapidly. Taken together, these results show that the generalized transport functions contain not only the hydrodynamic coefficients but also a nontrivial hierarchy of relaxation times and memory effects that becomes especially pronounced at finite density.

\section{Discussion and outlook}
\label{section-6}

In this work, we have developed an all-order linearized fluid/gravity description of a charged holographic plasma governed by five-dimensional Einstein-Maxwell-Chern-Simons theory. The construction is linear in the amplitude of the hydrodynamic perturbations but retains the complete dependence on frequency and spatial momentum. In the hydrodynamic regime, the perturbative expansion recovers the known shear, charge-diffusion, and chiral-vortical coefficients and provides the first frequency-dependent correction to the shear constitutive relation. Beyond this regime, the coupled radial equations were solved numerically without truncating the $\omega$- and $q$-dependence, yielding the generalized transport functions $\widetilde{\eta}$, $\widetilde{D}$, $\widetilde{\zeta}$, $\widetilde{\tau}$, $\widetilde{\xi}$, and $\widetilde{\lambda}_\Omega$  together with their associated memory kernels.

A central outcome of the analysis is that the derivative hierarchy of the constitutive relation does not coincide with the temporal hierarchy of the response. The leading shear and charge-diffusion functions, $\widetilde{\eta}$ and $\widetilde{D}$, exhibit a prompt response in the time domain. By contrast, the higher-gradient functions $\widetilde{\zeta}$ and $\widetilde{\tau}$ are strongly suppressed at early times and develop only after a finite buildup period. Most notably, the chiral-vortical function $\widetilde{\xi}$ also exhibits such a finite buildup time even though the corresponding vortical current appears already at first order in the derivative expansion. The temporal organization of transport therefore contains information that cannot be inferred from derivative counting alone. In this respect, the memory kernels provide a useful complement to the frequency-dependent constitutive coefficients by making the relaxation structure explicit.

The vortical current and vorticity-gradient tensor kernels both exhibit damped, sign-changing oscillations. Their qualitative similarity is consistent with the coupling
between the corresponding vector and tensor modes. Nevertheless, their temporal profiles are distinct: the tensor kernel reaches its principal extrema later and exhibits a weaker negative undershoot relative to its first positive maximum. The anomalous tensor response therefore contains information beyond an overall rescaling of the vortical current.

Finite density also affects the various transport channels in qualitatively different ways. The hydrodynamic shear viscosity retains its familiar universal behavior, while its finite-frequency and finite-momentum dependence becomes increasingly structured as the near-extremal regime is approached. The leading charge-diffusion coefficient, on the other hand, is progressively suppressed, whereas the charge-gradient tensor function $\widetilde{\tau}$ becomes increasingly prominent. Since $\widetilde{\tau}$ vanishes in the neutral background, this latter response is a genuinely finite-density higher-gradient effect. The leading and higher-gradient charge sectors therefore do not simply scale together with increasing density: the suppression of the leading diffusive response is accompanied by an enhancement of higher-gradient charge-induced transport. The anomalous channels also vanish at $M=1$, and their dominant spectral and temporal extrema become substantially more pronounced towards the near-extremal regime.

A common feature across the different channels is the increasing persistence of the late-time memory as the system approaches extremality. This occurs despite substantial differences in their short-time behavior. In particular, the diffusion kernel prompt response remains independent of $M$ while its late-time tail becomes longer-lived, showing that a reduction of the corresponding transport coefficient does not imply more rapid equilibration. Similarly, the higher-gradient and anomalous kernels acquire increasingly persistent tails toward the near-extremal regime. These observations are qualitatively consistent with the growing importance of the infrared dynamics associated with the near-horizon AdS$_2\times\mathbb{R}^3$ region of nearly extremal charged black branes ~\cite{Faulkner:2009wj,Edalati:2009bi,Edalati:2010hk,Davison:2013bxa}. A direct identification of the observed relaxation scales with specific AdS$_2$ excitations, however, would require an analysis of the complex-frequency spectrum and is beyond the scope of the present work.

The anomalous vortical channel provides an especially clear example of the additional information contained in the all-order response. The Chern-Simons interaction generates the chiral-vortical current already at leading derivative order, but its memory kernel builds up over a finite time and subsequently relaxes with pronounced sign-changing oscillatory structure characteristic of the higher-gradient shear channel. Increasing the charge parameter enhances the dominant extrema and the subsequent oscillatory tail, while the time at which the kernel reaches its maximum changes comparatively little. Thus, within the parameter range studied here, finite density modifies the strength of the anomalous response more strongly than the timescale over which it is established. Further, the influence of the Chern--Simons interaction extends beyond the explicitly vortical responses, generating corrections to the individual parity-even transport functions $\widetilde{\eta}$ and $\widetilde{\zeta}$. These corrections nevertheless cancel in the longitudinal combination $2\widetilde{\eta}-q^2\widetilde{\zeta}$, as established in Eq.~\eqref{eq:longitudinal_anomaly_cancellation}. Furthermore, the anomaly-dependent correction to $\widetilde{\eta}$ vanishes at zero spatial momentum,
consistent with the universal hydrodynamic ratio $\eta/s=1/(4\pi)$.

Spatial momentum provides a complementary probe of the nonlocal response. For all channels considered here, increasing $q$ generally reduces both the magnitude and the temporal extent of the corresponding memory kernel. The longest-lived response is therefore associated with long-wavelength perturbations, whereas shorter-wavelength fluctuations relax more rapidly. This common trend coexists with channel-dependent frequency structure and emphasizes that the all-order transport functions encode both temporal and spatial nonlocality.

Several limitations of the present analysis should be kept in mind. First, the construction is linear in the amplitude of the hydrodynamic perturbations and therefore does not capture nonlinear couplings between different modes. Such effects may become important for finite-amplitude disturbances, particularly at shorter wavelengths where gradient corrections are also expected to become significant. Second, varying the background parameter $M$ changes the charge density, temperature, and $\mu/T$ simultaneously. Consequently, the observed trends cannot be attributed solely to variations in charge density, but should instead be interpreted as reflecting the combined effects of these thermodynamic changes along the chosen family of charged black-brane backgrounds, particularly toward the near-extremal regime. Third, the exact extremal point \(M=3\) is excluded because the double zero of the blackening function changes the near-horizon radial problem. Furthermore, as extremality is approached, the conventional hydrodynamic description may become increasingly unreliable due to the emergence of additional long-lived excitations and enhanced infrared dynamics~\cite{He:2021jna}. Nevertheless, our results exhibit increasingly persistent memory kernels in this regime, highlighting the importance of long-time dynamics near extremality. Finally, the present analysis is restricted to the real-frequency axis. An investigation of the analytic structure of the transport functions in the complex-frequency plane, including their poles and possible branch cuts, would provide further insight into the microscopic relaxation modes and long-time dynamics underlying the observed memory kernels.

These observations suggest several natural extensions. It would be particularly useful to determine the quasinormal-mode spectrum of the coupled Einstein-Maxwell-Chern-Simons system and relate its low-lying modes directly to the characteristic timescales extracted from the memory kernels. The exact extremal geometry deserves a separate treatment, where the AdS$_2$ infrared sector may be isolated analytically. Extending the calculation beyond linear order would allow one to study mode coupling and nonlinear anomalous transport, while the inclusion of external electromagnetic fields would provide access to a broader class of anomalous magnetohydrodynamic phenomena~\cite{Jimenez-Alba:2014iia,Lublinsky:2026wvt}.

In summary, the all-order fluid/gravity construction provides a unified description of momentum, charge, higher-gradient, and anomalous transports in a finite-density strongly coupled plasma. The resulting generalized transport functions reproduce the hydrodynamic coefficients at small frequency and momentum while retaining the nonlocal response away from this limit. Their memory kernels reveal a hierarchy of prompt and initially suppressed responses, increasingly long-lived relaxation toward extremality, and a clear separation between derivative order and temporal onset. These results demonstrate that the real-time structure of transport contains information that is not visible in a finite-order hydrodynamic expansion and provide a basis for further studies of infrared relaxation and anomalous dynamics in charged holographic systems.

\appendix
\section{Linearized bulk equations before tensor decomposition}
\label{app:linearized-bulk-equations}
In this appendix, we collect the component form of the linearized Einstein-Maxwell-Chern-Simons equations prior to the scalar, vector, and tensor decomposition introduced in Sec.~\ref{section-2}. These equations provide the starting point for the reduced radial systems \eqref{eq:39}-\eqref{eq:44} used in the main text.

We denote the Einstein and Maxwell equations schematically by $E_{AB}=0$ and $\mathcal{M}_{A}=0$, respectively. Throughout this appendix,
\begin{equation}
\partial\beta \equiv \partial_i\beta_i,
\qquad
\partial j \equiv \partial_i j_i,
\qquad
\partial g \equiv \partial_i g_i,
\qquad
\partial^2 \equiv \partial_i\partial_i .
\end{equation}
The correction functions $\{h,k,w,j_i,g_i,\alpha_{ij}\}$ are those defined in Sec.~\ref{section-2}.

\subsection{Scalar sector}

The simplest scalar Einstein equation is the $rr$ component,
\begin{equation}
\label{app:eq:Err}
5\partial_r h+r\partial_r^2 h=0 .
\end{equation}
As discussed in the main text, horizon regularity, boundary normalizability, and the Landau-frame condition imply
\begin{equation}
h=0 .
\end{equation}

The scalar Maxwell equation $\mathcal{M}_v=0$ then gives
\begin{equation}
\label{app:eq:Mv}
\partial_r w
-r\partial_r^2 w
+5r^5\partial g
+r^6\partial_r\partial g
=0 .
\end{equation}

The $rv$ component of the Einstein equations determines the radial derivative of the remaining scalar function $k$:
\begin{equation}
\label{app:eq:Erv}
\begin{aligned}
 3r^2 \partial_r k {}=& \frac{12}{r} q_0 w - 6 q_0 \partial_r w + 6r^4 \partial \beta + r^3 \partial_v \partial \beta - 36 r^4 \partial j - 6 r^5 \partial_r \partial j - r^3 \partial_i \partial_j \alpha_{ij}.
\end{aligned}
\end{equation}
Since $w$, $j_i$, and $\alpha_{ij}$ are determined by the remaining sectors, Eq.~\eqref{app:eq:Erv} is most conveniently integrated only after the vector and tensor equations have been solved.

\subsection{Vector sector}

The vector sector is governed by the $ri$ and $vi$ components of the Einstein equations together with the spatial Maxwell equation $\mathcal{M}_i=0$. One linear combination of the Einstein equations is a constraint equation, while the independent dynamical combination may be represented by $E_{ri}=0$.

The latter takes the form
\begin{equation}
\label{app:eq:Eri}
\begin{aligned}
-6r^4\partial_r^2j_i
={}&
\left(
\partial^2\beta_i-\partial_i\partial\beta
\right)
+3r\partial_v\beta_i
+30r^3q_0g_i
+
6r^4q_0\partial_rg_i
+30r^3\partial_rj_i
+r^2\partial_r\partial_j\alpha_{ij}.
\end{aligned}
\end{equation}

The spatial Maxwell equation is
\begin{equation}
\label{app:eq:Mi}
\begin{aligned}
 0{}= & \left(q_0^2 r^2 - m_0 r^4 + r^8\right)\partial_r^2 g_i + \left(7q_0^2 r - 9m_0 r^3 + 13r^7\right)\partial_r g_i + 5\left(q_0^2 - 3m_0 r^2 + 7r^6\right)g_i \\
        & + 12q_0 \partial_r j_i + 11r^5 \partial_v g_i + 2r^6 \partial_r \partial_v g_i - r^4\left(\partial_i \partial g - \partial^2 g_i\right) 
        + 8\sqrt{3}q_0 r^2 \kappa \epsilon_{ijk} \partial_j g_k \\
        & + \frac{1}{r}\partial_r \partial_i w - \frac{1}{r^2}\partial_i w
        - \frac{1}{r^5}\left[-8\sqrt{3}\kappa q_0^2 \ell_i + r^3 \partial_i q^{(0)} + r^3 q_0 \partial_v \beta_i + q_0 r^2 (\partial_i \partial \beta - \partial^2 \beta_i)\right].
\end{aligned}
\end{equation}

\subsection{Tensor sector}

The independent tensor equation is obtained by taking the symmetric traceless projection of the spatial Einstein equations $E_{ij}=0$. It reads
\begin{equation}
\label{app:eq:Eij}
\begin{aligned}
0={}&
r\left(r^6-m_0r^2+q_0^2\right)
\partial_r^2\alpha_{ij}
-
\left(
-5r^6+m_0r^2+q_0^2
\right)
\partial_r\alpha_{ij}
+
3r^4\partial_v\alpha_{ij}
+
2r^5\partial_v\partial_r\alpha_{ij}
\\
&+
r^3
\Bigg[
\partial^2\alpha_{ij}
-
\left(
\partial_i\partial_k\alpha_{jk}
+
\partial_j\partial_k\alpha_{ik}
-
\frac{2}{3}
\delta_{ij}\partial_k\partial_l\alpha_{kl}
\right)
\Bigg]
-
18r^4
\left(
\partial_i j_j
+
\partial_j j_i
-
\frac{2}{3}\delta_{ij}\partial j
\right)
\\
&-
6r^5\partial_r
\left(
\partial_i j_j
+
\partial_j j_i
-
\frac{2}{3}\delta_{ij}\partial j
\right)
+
3r^4
\left(
\partial_i\beta_j
+
\partial_j\beta_i
-
\frac{2}{3}\delta_{ij}\partial\beta
\right)
\\
&+
r^3\partial_v
\left(
\partial_i\beta_j
+
\partial_j\beta_i
-
\frac{2}{3}\delta_{ij}\partial\beta
\right).
\end{aligned}
\end{equation}

Equations~\eqref{app:eq:Mv}-\eqref{app:eq:Eij} are linear in the hydrodynamic perturbations and their boundary derivatives. Fourier transforming according to
\begin{equation}
\partial_\mu
\longrightarrow
(-i\omega,i\vec q)
\end{equation}
and substituting the scalar, vector, and tensor decomposition \eqref{eq:37} reduce these partial differential equations to the radial systems \eqref{eq:39}-\eqref{eq:43}. The scalar equation \eqref{app:eq:Erv} subsequently yields Eq.~\eqref{eq:44} for $k$.

\section{Perturbative solutions of the radial equations}
\label{app:perturbative-radial}

In this appendix, we provide the intermediate radial equations and explicit perturbative solutions underlying the hydrodynamic expansion presented in Sec.~\ref{section-3}. The coefficient functions are expanded according to Eq.~\eqref{eq:gradient-expansion}, with the understanding that the total derivative order also includes the explicit derivatives carried by the tensor structures in Eq.~\eqref{eq:37}. Throughout this appendix, the same horizon regularity, boundary normalizability, and Landau-frame conditions introduced in Sec.~\ref{section-2} are imposed.

\subsection{Zeroth-order solution}
\label{app:zeroth-order}

At zeroth order in the expansion parameter $\lambda$, only $b_1^{(0)}$ and $V_1^{(0)}$ contribute. The corresponding Sector-I equations are
\begin{equation}
\label{app:eq:zero-system}
\begin{aligned}
0={}&
r^4\partial_r^2 b_1^{(0)}
+5r^3\partial_r b_1^{(0)}
+r^4q_0\partial_r V_1^{(0)}
+5r^3q_0V_1^{(0)},
\\
0={}&
\left(q_0^2r^2-m_0r^4+r^8\right)
\partial_r^2V_1^{(0)}
+
\left(
7q_0^2r-9m_0r^3+13r^7
\right)
\partial_rV_1^{(0)}
\\
&+
5\left(
q_0^2-3m_0r^2+7r^6
\right)V_1^{(0)}
+
12q_0\partial_rb_1^{(0)}.
\end{aligned}
\end{equation}
The first equation integrates once to
\begin{equation}
\label{app:eq:zero-first-integral}
r^5
\left(
\partial_rb_1^{(0)}
+
q_0V_1^{(0)}
\right)
=C .
\end{equation}
The boundary conditions require $C=0$, and hence
\begin{equation}
\label{app:eq:zero-V1}
V_1^{(0)}
=
-\frac{1}{q_0}
\partial_rb_1^{(0)} .
\end{equation}
Substitution into the second equation in Eq.~\eqref{app:eq:zero-system} gives a homogeneous equation for $b_1^{(0)}$. The solution compatible with the boundary and frame
conditions is
\begin{equation}
\label{app:eq:zero-solution}
b_1^{(0)}=0,
\qquad
V_1^{(0)}=0,
\end{equation}
which is the result quoted in Eq.~\eqref{eq:zero-order}.

\subsection{First-order radial systems}
\label{app:first-order-systems}

At first derivative order, the nontrivial perturbations separate into three sectors associated with shear transport, charge diffusion, and the Chern-Simons-induced vortical response.

\subsubsection{Shear sector}

The relevant radial equations are
\begin{equation}
\label{app:eq:first-shear-system}
\begin{aligned}
0={}&
r\left(r^6-m_0r^2+q_0^2\right)
\partial_r^2s_1^{(0)}
-
\left(
-5r^6+m_0r^2+q_0^2
\right)
\partial_rs_1^{(0)}
+3r^4,
\\[1mm]
0={}&
6r^4\partial_r^2b_1^{(1)}
+
30r^3\partial_rb_1^{(1)}
+
6r^4q_0\partial_rV_1^{(1)}
+
30r^3q_0V_1^{(1)}
-3i\omega r,
\\[1mm]
0={}&
\left(
q_0^2r^2-m_0r^4+r^8
\right)
\partial_r^2V_1^{(1)}
+
\left(
7q_0^2r-9m_0r^3+13r^7
\right)
\partial_rV_1^{(1)}
\\
&+
5\left(
q_0^2-3m_0r^2+7r^6
\right)V_1^{(1)}
+
12q_0\partial_rb_1^{(1)}
+
\frac{i\omega q_0}{r^2}.
\end{aligned}
\end{equation}

The tensor equation can equivalently be written as
\begin{equation}
\label{app:eq:s10-total}
r^2
\partial_r
\left[
r^5f(r)
\partial_rs_1^{(0)}
\right]
+
3r^4
=0 .
\end{equation}
The above equation can be solved by imposing physical boundary conditions to get
\begin{equation}\label{app:eq:s10-integral}
    s_1^{(0)}(\rho)=\frac{1}{R}\int^\infty_\rho \frac{p(p^2+p+1)}{(p+1)(p^4+p^2-M+1)}dp
\end{equation}
in rescaled variables. Near the conformal boundary,
\begin{equation}\label{app:eq:s10-asymptotic}
s_1^{(0)}(r)=\frac{1}{r}-\frac{R^3}{4r^4}+\mathcal{O}\left(r^{-5}\right) .
\end{equation}

For the vector modes, the second equation of
Eq.~\eqref{app:eq:first-shear-system} can be written as
\begin{equation}
\label{app:eq:b11-integrated}
\frac{6}{r}
\partial_r
\left[
r^5
\left(
\partial_rb_1^{(1)}
+
q_0V_1^{(1)}
\right)
\right]
-
3i\omega r
=0,
\end{equation}
which gives
\begin{equation}
\label{app:eq:V11-b11}
V_1^{(1)}
=
\frac{i\omega}{6q_0r^2}
-
\frac{1}{q_0}
\partial_rb_1^{(1)} .
\end{equation}
Substitution into the remaining vector equation yields
\begin{equation}
\label{app:eq:b11-decoupled}
\partial_r
\left[
\frac{1}{r}
\partial_r
\left(
r^7f(r)^2
\partial_r
\left[
\frac{b_1^{(1)}}{f(r)}
\right]
\right)
\right]
=
\frac{
i\omega
\left(
q_0^2-m_0r^2+5r^6
\right)
}{
2r^4
}.
\end{equation}
The resulting solution is
\begin{equation}
\label{app:eq:first-shear-vector-solution}
b_1^{(1)}(\omega,q,r)=-\frac{i\omega}{6r}+\frac{i\omega q_0^2(1+M)}{8MRr^6}+\mathcal{O}(r^{-7}),
\qquad
V_1^{(1)}(\omega,q,r)=\frac{3i\omega q_0(1+M)}{4MRr^7}+\mathcal{O}(r^{-8}).
\end{equation}

\subsubsection{Charge-diffusion sector}

The diffusion-sector equations are
\begin{equation}
\label{app:eq:first-diffusion-system}
\begin{aligned}
0={}&
r^4\partial_r^2b_3^{(0)}
+
5r^3\partial_rb_3^{(0)}
+
r^4q_0\partial_rV_3^{(0)}
+
5r^3q_0V_3^{(0)},
\\
0={}&
\left(
q_0^2r^2-m_0r^4+r^8
\right)
\partial_r^2V_3^{(0)}
+
\left(
7q_0^2r-9m_0r^3+13r^7
\right)
\partial_rV_3^{(0)}
\\
&+
5\left(
q_0^2-3m_0r^2+7r^6
\right)V_3^{(0)}
+
12q_0\partial_rb_3^{(0)}
-
\frac{1}{r^2}.
\end{aligned}
\end{equation}
In terms of the dimensionless coordinate $\rho=r/R$, the first equation
implies
\begin{equation}
\label{app:eq:V30-b30}
V_3^{(0)}
=
-\frac{1}{R^4Q}
\partial_\rho b_3^{(0)} .
\end{equation}
Substituting Eq.~\eqref{app:eq:V30-b30} into Eq.~\eqref{app:eq:first-diffusion-system} and simplifying, we obtain a single third-order equation for $b_3^{(0)}$,
\begin{equation}
\partial_\rho\left(\frac{1}{\rho}\partial_\rho\left[\rho^7 f(\rho)^2\partial_\rho\left(\frac{b_3^{(0)}}{f(\rho)}\right)\right]\right)=-\frac{Q}{R^4 \rho^4}.
\end{equation}
The solution for $b_3^{(0)}$ satisfying the boundary condition is
\begin{equation}\label{app:eq:b30}
b_3^{(0)}=\frac{Q}{3R^4}f(\rho)\int_\rho^\infty\frac{dp}{f(p)^2}\left(\frac{1}{p^8}-\frac{3}{4p^7}\left(1+\frac{1}{M}\right)\right)
\end{equation}

Expanding the solution near the conformal boundary, $r\to\infty$, we find that the solution quoted in the main text is
\begin{equation}
\label{app:eq:first-diffusion-solution}
\begin{aligned}
b_3^{(0)}&=-\frac{Q(1+M)}{24MR^4\rho^6}+\mathcal{O}(\rho^{-7}),
\qquad
V_3^{(0)}&=-\frac{1+M}{4MR^8\rho^7}+\mathcal{O}(\rho^{-8}).
\end{aligned}
\end{equation}

\subsubsection{Chern-Simons sector}

The vorticity sector obeys
\begin{equation}
\label{app:eq:first-CS-system}
\begin{aligned}
{} 0= & \
        r^4\partial_r^2 b_4^{(0)}+5r^3\partial_r b_4^{(0)}+r^4q_0\partial_r V_4^{(0)}+5r^3q_0V_4^{(0)} \\ 0= & \
         \left(q_0^2 r^2-m_0r^4+r^8\right)\partial_r^2 V_4^{(0)}+\left(7q_0^2r-9m_0r^3+13r^7\right)\partial_r V_4^{(0)}+5\left(q_0^2-3m_0r^2+7r^6\right)V_4^{(0)} \\ & +12q_0\partial_r b_4^{(0)}+\frac{8\sqrt{3}\kappa q_0^2}{r^5} .
\end{aligned}
\end{equation}
Eliminating $V_4^{(0)}$ gives
\begin{equation}
\label{app:eq:b40-decoupled}
\partial_\rho
\left\{
\frac{1}{\rho}
\partial_\rho
\left[
\rho^7 f(\rho)^2
\partial_\rho
\left(
\frac{b_4^{(0)}}{f(\rho)}
\right)
\right]
\right\}
=
\frac{8\sqrt{3}Q^3\kappa}{R\rho^7}.
\end{equation}
A convenient integral representation of the solution is
\begin{equation}
\label{app:eq:b40-integral}
b_4^{(0)}=\frac{\sqrt{3}Q^3\kappa}{3R}f(\rho)\int_\rho^\infty\frac{dp}{f(p)^2}\left(\frac{3}{Mp^7}-\frac{1}{p^{11}}\right),
\end{equation}
where the integration constants are fixed by horizon regularity and boundary normalizability. Evaluating the integral gives the exact first-order radial profiles
\begin{equation}
\label{app:eq:first-CS-asymptotic}
\begin{aligned}
b_4^{(0)}
&=
\frac{\sqrt{3}\kappa q_0^3}{6m_0 r^6},
\qquad
V_4^{(0)}
=
\frac{\sqrt{3}\kappa q_0^2}
{m_0 r^7}.
\end{aligned}
\end{equation}

\subsubsection{Scalar sector}

Once the vector and tensor sectors have been determined, the remaining scalar equation reduces at first derivative order to
\begin{equation}
\label{app:eq:k1-radial}
\partial_rk^{(1)}
=
2r^2\partial_i\beta_i .
\end{equation}
Its solution is
\begin{equation}
\label{app:eq:k1}
k^{(1)}
=
\frac{2}{3}r^3
\partial_i\beta_i .
\end{equation}

\subsection{Complete first-order bulk correction}
\label{app:first-order-bulk-fields}

Combining the solutions above, the first-order metric and gauge-field corrections take the form
\begin{equation}
\label{app:eq:first-complete}
\begin{aligned}
{} & \delta^{ij}g_{ij}^{(1)}(r)=0,\\ &
       g_{vv}^{(1)}(r)=\frac{2}{3}r \partial_m \beta_m,\\ &
       g_{vr}^{(1)}(r)=0,\\ &
       g_{vi}^{(1)}(r)=r \partial_v \beta_i+\frac{\sqrt{3}q_0^3\kappa}{m_0r^4}\ell_i-\frac{q_0(1+M)}{4MRr^4}\left(\partial_i q^{(0)}+3q_0\partial_v\beta_i\right),\\ &
       g_{ij}^{(1)}(r)=2r^2\left(\frac{1}{r}-\frac{R^3}{4r^4}\right)\sigma_{ij},\\ &
        A_v^{(1)}(r)=0,\\ &
       A_i^{(1)}(r)=\frac{3\kappa q_0^2}{2m_0r^2}\ell_i-\frac{\sqrt{3}(1+M)}{8MR r^2}\left(\partial_iq^{(0)}+3q_0\partial_v\beta_i\right).
\end{aligned}
\end{equation}
The normalizable parts of Eq.~\eqref{app:eq:first-complete}, inserted into the holographic expressions for the stress tensor and current, give Eqs.~\eqref{eq:first-stress}-\eqref{eq:first-coefficients} of the main text.

\subsection{Second-order radial systems}\label{app:second-order-radial-systems}

\subsubsection{Second-order shear sector}
\label{app:second-order-shear}

At the next order, the Sector-I equations relevant for the shear channel are
\begin{equation}
\label{app:eq:second-shear-system}
\begin{aligned}
0={}&
6r^4\partial_r^2b_1^{(2)}
+
30r^3\partial_rb_1^{(2)}
+
6r^4q_0\partial_rV_1^{(2)}
+
30r^3q_0V_1^{(2)}-q^2r^2\partial_r s_1^{(0)}-q^2,
\\[1mm]
0={}&
\left(
q_0^2r^2-m_0r^4+r^8
\right)
\partial_r^2V_1^{(2)}
+
\left(
7q_0^2r-9m_0r^3+13r^7
\right)
\partial_rV_1^{(2)}
\\
&+
5\left(
q_0^2-3m_0r^2+7r^6
\right)V_1^{(2)}
+
12q_0\partial_rb_1^{(2)}
-
11i\omega r^5V_1^{(1)}
\\
&-
2i\omega r^6\,\partial_rV_1^{(1)}
+8\sqrt{3}\kappa q_0r^2q^2V_4^{(0)}-q^2 \frac{q_0}{r^3},
\\[1mm]
0={}&
r\left(
r^6-m_0r^2+q_0^2
\right)
\partial_r^2s_1^{(1)}
-
\left(
-5r^6+m_0r^2+q_0^2
\right)
\partial_rs_1^{(1)}-
2i\omega r^5
\partial_rs_1^{(0)}
\\
&
-
3i\omega r^4s_1^{(0)}
-
18r^4b_1^{(1)}
-
6r^5\partial_rb_1^{(1)}-i\omega r^3 .
\end{aligned}
\end{equation}
Using the lower-order solutions, the tensor equation reduces to
\begin{equation}
\label{app:eq:second-shear-reduced}
\partial_\rho
\left[
\rho^5f(\rho)
\partial_\rho s_1^{(1)}
\right]
=
\frac{i\omega}{R^2}\bigg[2\rho^3F'(\rho)+3\rho^2F(\rho)-\rho-18(M-1)\partial_\rho\left(\rho^3 B(\rho)\right)\bigg],
\end{equation}
with
\begin{equation}
\label{app:eq:F2}
\begin{aligned}
    & {} F(\rho)=\int_\rho^\infty\left(\frac{p(p^2+p+1)}{(p+1)(p^4+p^2-M+1)}\right)dp,\\ &
    B(\rho)=\frac{1}{3}f(\rho)\int_\rho^\infty\frac{dp}{f(p)^2}\left(\frac{1}{p^8}-\frac{3}{4p^7}\left(1+\frac{1}{M}\right)\right).
\end{aligned}
\end{equation}
An explicit integral representation is
\begin{equation}
\label{app:eq:s11-integral}
\begin{aligned}
s_1^{(1)}(\rho)
={}&
\int_\rho^\infty
\frac{dx}{x^5f(x)}
\int_1^xdy\,
\Bigg[-
\frac{i\omega}{R^2}
\bigg(
2y^3F'(y)+3y^2F(y)-y-18(M-1)\partial_y\left(y^3B(y)
\right)\bigg)
\Bigg].
\end{aligned}
\end{equation}
Near the asymptotic boundary,
\begin{equation}
\label{app:eq:s11-asymptotic}
s_1^{(1)}(r)
=
-\frac{i\omega R^2}{8r^4}\left(2+\frac{M}{\sqrt{4M-3}}\log\left(\frac{3-\sqrt{4M-3}}
{3+\sqrt{4M-3}}\right)-\frac{3(M-1)}{2M}\right)+\mathcal{O}(r^{-5}).
\end{equation}
\subsubsection{Second-order diffusive sector}
Further, at second-order the charge sector (Sector-III) develops a tensorial transport contribution whose relevant equation is
\begin{equation}\label{app:eq:diff_tensor}
        r(r^6-m_0r^2+q_0^2)\partial_r^2 s_3^{(0)} - (-5r^6+m_0r^2+q_0^2)\partial_r s_3^{(0)}  - 36r^4 b_3^{(0)} - 12 r^5 \partial_r b_3^{(0)} = 0
\end{equation}
Rewriting Eq.~\eqref{app:eq:diff_tensor} in terms of the dimensionless radial coordinate $\rho$ and the dimensionless parameter $M$, we obtain
\begin{equation}\label{eq:85b}
    \partial_\rho \left(\rho^5 f(\rho) \partial_\rho s_3^{(0)}\right)=\frac{12Q}{R^5} \partial_\rho\left(\rho^3 B(\rho)\right)
\end{equation}
where $B(\rho)$ is used as defined in Eq.~\eqref{app:eq:F2} and we have used the rescaling $Q^2=M-1$.

Consequently, the near-boundary behavior of the second-order tensor transport function is
\begin{equation}\label{app:eq:s30}
\begin{aligned}
    s_3^{(0)} {} &  =-\frac{12Q}{R^5}\int_\rho^\infty d\zeta\left(\frac{\zeta}{(M-1)-M\zeta^2+\zeta^6}\right) \int_1^\zeta d\xi\left( \partial_\xi\left[\xi^3 B(\xi)\right]\right), \\ &
    = -\frac{\sqrt{M-1}}{8MR^5\rho^4}+\mathcal{O}(\rho^{-7}).
    \end{aligned}
\end{equation}

\subsubsection{Second-order vorticity sector}
Similarly, the vorticity sector also develops a tensorial transport coupled to the vorticity vector and its leading equation is
\begin{equation}\label{eq:vor_tensor}
    r(r^6-m_0r^2+q_0^2)\partial_r^2s_4^{(0)}+(5r^6-m_0r^2-q_0^2)\partial_rs_4^{(0)}-18r^4b_4^{(0)}-6r^5\partial_rb_4^{(0)}=0
\end{equation}
Rewriting~\eqref{eq:vor_tensor}, in terms of dimensionless quantities, we obtain
\begin{equation}
    \pa_\rho\left(\rho^5f(\rho)\pa_\rho s_4^{(0)}\right)=-\frac{3\sqrt{3}\kappa Q^3}{MR^2\rho^4}
\end{equation}
Imposing and boundary condition and regularity at $\rho=1$, we obtain,
\begin{equation}
    s_4^{(0)}(\rho)=\frac{\sqrt{3}\kappa Q^3}{MR^2}\int_\rho^\infty\frac{p^2+p+1}{p^2(p+1)(p^4+p^2-M+1)}dp
\end{equation}
Whose near conformal boundary solution is,
\begin{equation}\label{app:eq:s40-asymptotic}
    s_4^{(0)}(\rho)=\frac{\sqrt{3}\kappa (M-1)^{3/2}}{4MR^2\rho^4}+\mathcal{O}(\rho^{-7})
\end{equation}
Insertion of these normalizable coefficients~\eqref{app:eq:s11-asymptotic},~\eqref{app:eq:s30}  and~\eqref{app:eq:s40-asymptotic} into the holographic stress tensor gives Eq.~\eqref{eq:second-stress} in the main text.

\subsection{Third-order radial systems}\label{app:third-order}
The leading coefficient $s_2^{(0)}$ is sourced by the second-order vector profiles $b_2^{(0)}$ and $V_2^{(0)}$. We first solve their coupled radial equations using the previously determined first-order shear and vortical profiles, and then integrate the tensor equation to extract $\zeta$ from its normalizable boundary coefficient. The relevant equations are:
\begin{equation}\label{eq:third_tensor}
    \begin{aligned}
        {} 0= & \ 6r^4 \partial_r^2b_2^{(0)} + 30r^3\partial_r b_2^{(0)} + 6r^4 q_0\partial_r V_2^{(0)} + 30r^3 q_0 V_2^{(0)} + \frac{1}{3}r^2 \partial_r s_1^{(0)} - 1  \\[1mm] 0 = & \
        \left(q_0^2r^2-m_0r^4+r^8\right)\partial_r^2 V_2^{(0)} + \left(7q_0^2 r-9m_0r^3+13r^7\right)\partial_r V_2^{(0)} + 5\left(q_0^2-3m_0r^2+7r^6\right)V_2^{(0)} \\ & + 12q_0\partial_r b_2^{(0)}  +8\sqrt{3}\kappa q_0 r^2 V_4^{(0)} - \frac{q_0}{r^3} \\[1mm]  0 = & \ 
        r\left(r^6-m_0r^2+q_0^2\right)\partial_r^2 s_2^{(0)} - \left(-5r^6+m_0r^2+q_0^2\right)\partial_r s_2^{(0)} - 36r^4 b_2^{(0)} - 12 r^5 \partial_r b_2^{(0)} - \frac{2}{3}r^3 s_1^{(0)}
    \end{aligned}
\end{equation}

We use the $V_4^{(0)}$ obtained in Eq.~\eqref{app:eq:first-CS-asymptotic}, we obtain
\begin{equation}
    \frac{6}{\rho}\pa_\rho\left[\rho^5\left(R^2\pa_\rho b_2^{(0)}+QR^6V_2^{(0)}\right)\right]+\frac{1}{3}\rho^2 F'(\rho)-1=0
\end{equation}

Which implies
\begin{equation}\label{eq:third_v2}
    V_2^{(0)}(\rho)=\frac{1}{QR^6}\left[\frac{1}{\rho^5}\left(\frac{1}{18}\int^\infty_\rho\left(p^3F'(p)+p\right)dp+\frac{\rho^2}{9}\right)-R^2\pa_\rho b_2^{(0)}\right]
\end{equation}
Substituting \eqref{eq:third_v2} in \eqref{eq:third_tensor} and solving by imposing regularity at $\rho=1$ and boundary conditions we get
\begin{equation}
    b_2^{(0)}(\rho)=-\frac{f(\rho)}{R^2}\int_\rho^\infty \frac{dp}{p^7f(p)^2}\left(\frac{p^4}{9}+\frac{p}{18}-\frac{Q^2}{6p^2}+\frac{\kappa^2Q^4}{Mp^4}+\frac{2M^2-3M+3}{24M}-\frac{3\kappa^2Q^4}{M^2}\right)
\end{equation}
Whose near conformal boundary expansion is
\begin{equation}
    b_2^{(0)}(\rho)=\frac{1}{R^2}\left(-\frac{1}{18\rho^2}-\frac{1}{90\rho^5}+\frac{1}{\rho^6}\left[\frac{M}{216}+\frac{Q^2}{48M}+\frac{\kappa^2Q^4}{2M^2}\right]\right)+\mathcal{O}(\rho^{-8})
\end{equation}
and $V_2^{(0)}$ behaves as
\begin{equation}
    V_2^{(0)}(\rho)=\frac{1}{R^6\rho^7}\left(\frac{Q}{8M}+\frac{3\kappa^2Q^3}{M^2}\right)
\end{equation}
Now, we use Eq.~\eqref{eq:third_tensor} to calculate $s_2^{(0)}$ by rewriting it as
\begin{equation}
    R^3\pa_\rho\left(\rho^5f(\rho)\pa_\rho s_2^{(0)}(\rho)\right)=12R^2\pa_\rho\left(\rho^3b_2^{(0)}\right)+\frac{2}{3}\rho F(\rho)
\end{equation}
Whose solution using horizon regularity and boundary conditions is,
\begin{equation}
    s_2^{(0)}(\rho)=-\frac{1}{R^3}\int_\rho^\infty\frac{dx}{x^5f(x)}\int_1^x d\xi\left(12R^2\pa_\xi\left(\xi^3b_2^{(0)}(\xi)\right)+\frac{2}{3}\xi F(\xi)\right)
\end{equation}
Whose near conformal boundary solution is
\begin{equation}
\begin{aligned}\label{app:eq:s20-asymptotic}
s_2^{(0)}(r)={}&-\frac{R}{4r^4}\Bigg\{\frac{2M+3}{12M}-\frac{6\kappa^2(M-1)^2}{M^2}-\frac{\left(1+\sqrt{4M-3}\right)^{3/2}}
{6\sqrt{2}\sqrt{4M-3}}\arctan\!\sqrt{\frac{1+\sqrt{4M-3}}{2}}\\&+\frac{\left(\sqrt{4M-3}-1\right)^{3/2}}{6\sqrt{2}\sqrt{4M-3}}\operatorname{arctanh}\!\sqrt{\frac{\sqrt{4M-3}-1}{2}}-\frac{1}{6\sqrt{4M-3}}\log\!\left(\frac{3+\sqrt{4M-3}}{3-\sqrt{4M-3}}\right)\Bigg\}\\&+\mathcal{O}(r^{-6}).
\end{aligned}
\end{equation}

Insertion of this normalizable coefficient~\eqref{app:eq:s20-asymptotic} into the holographic stress tensor gives Eq.~\eqref{eq:third_order_stress} in the main text.

\section*{Acknowledgements}
K.R. expresses gratitude to R. Loganayagam for fruitful discussions and valuable insights, and to Sanhita Parihar for helpful discussions. A.N.K. thanks Michael Lublinsky for insightful discussions on the all-order hydrodynamic formulation. A.N.K. also acknowledges financial support from the Croatian Science Foundation under Project No. IP-2025-02-8625, \emph{Quantum Aspects of Gravity.}

\bibliography{mybib}
\end{document}